\documentclass{article}
\PassOptionsToPackage{square,comma,numbers,sort&compress}{natbib}  
\usepackage[preprint]{neurips_2020}
\usepackage[breaklinks,colorlinks,
linkcolor=red,
anchorcolor=blue,
citecolor=blue
]{hyperref}
\usepackage[utf8]{inputenc} 
\usepackage[T1]{fontenc}   
\usepackage{url}          
\usepackage{booktabs}
\usepackage{amsfonts} 
\usepackage{nicefrac}
\usepackage{microtype}
\usepackage{amsmath}
\usepackage{amsfonts}
\usepackage{amssymb}
\usepackage{mathtools}
\usepackage{amsthm}
\usepackage{appendix}
\usepackage{subcaption}
\usepackage{url}
\usepackage{enumitem}
\usepackage{bm}
\usepackage{graphicx}
\graphicspath{ {figures/} }

\usepackage{kantlipsum, wrapfig,graphicx}

\newtheorem{theorem}{Theorem}

\newtheorem{lemma}{Lemma}[section]
\newtheorem{definition}{Definition}[section]

\DeclarePairedDelimiter{\norm}{\lVert}{\rVert}

\DeclareMathOperator{\relu}{ReLU}

\title{Neural Certificates for Safe Control Policies}

\author{Wanxin Jin\\
	Purdue University\\
	wanxinjin@gmail.com
	\hspace{-1.9mm}
	\And
	Zhaoran Wang \\
	Northwestern University\\
	zhaoranwang@gmail.com
	\hspace{-1.9mm}
	\And
	Zhuoran Yang \\
	Princeton University\\
	zy6@princeton.edu
	\hspace{-1.9mm}
	\And
	Shaoshuai Mou \\
	Purdue University\\
	mous@purdue.edu}

\begin{document}

\maketitle

\begin{abstract}
	This paper develops an approach to learn a policy of a dynamical system that is  guaranteed to be both provably safe and goal-reaching. Here, the safety means that a policy must not drive  the state of the system to any unsafe region, while the  goal-reaching requires the trajectory of the controlled system  asymptotically converges to a goal region (a generalization of stability). We obtain the safe and goal-reaching policy by jointly learning two additional certificate functions: a barrier function that guarantees the safety and a developed Lyapunov-like function to fulfill the goal-reaching requirement, both of which are represented by neural networks. We show the effectiveness of the method to learn both safe and goal-reaching policies on various systems, including pendulums, cart-poles, and UAVs.

\end{abstract}

\section{Introduction}
We summarize three levels of priorities in order for policy learning tasks:
\begin{align}
&\textbf{Safety:}\quad \emph{The motion of a system under a policy must not run into any unsafe region.} \nonumber\\
&\textbf{Stability:} \quad\emph{The motion of a system under a policy is bounded around or attractive to a point. }\nonumber\\
&\textbf{Optimality:} \quad \emph{The trajectory of a system under a policy maximizes an accumulative reward. }\nonumber
\end{align}
Existing techniques extensively focus on learning a policy of optimality. These methods typically use the framework of reinforcement learning \cite{sutton2018reinforcement}, which evaluates and improves the policy by interacting with environments/systems. Although achieving notable progress \cite{oh2016control,mnih2013playing,mnih2015human}, their deployments are still limited and a key challenge is the safety issue during  the exploration process \cite{amodei2016concrete}. 
In practice, safety and stability always come first before the pursuit of  policy optimality. Although recent work \cite{kolter2019learning,chang2019neural,richards2018lyapunov,chow2018lyapunov} began to consider safety in the development of learning techniques, most of them use Lyapunov theory \cite{lyapunov1892general} and achieve safety by requiring  attractiveness, i.e., stability, of the system's behavior. Such attractiveness requirement, although sufficient to guarantee safety, is typically conservative. 

\begin{wrapfigure}[14]{r}{0pt}
	\raisebox{0pt}[\dimexpr\height-1.2\baselineskip]{\includegraphics[width=0.25\textwidth]{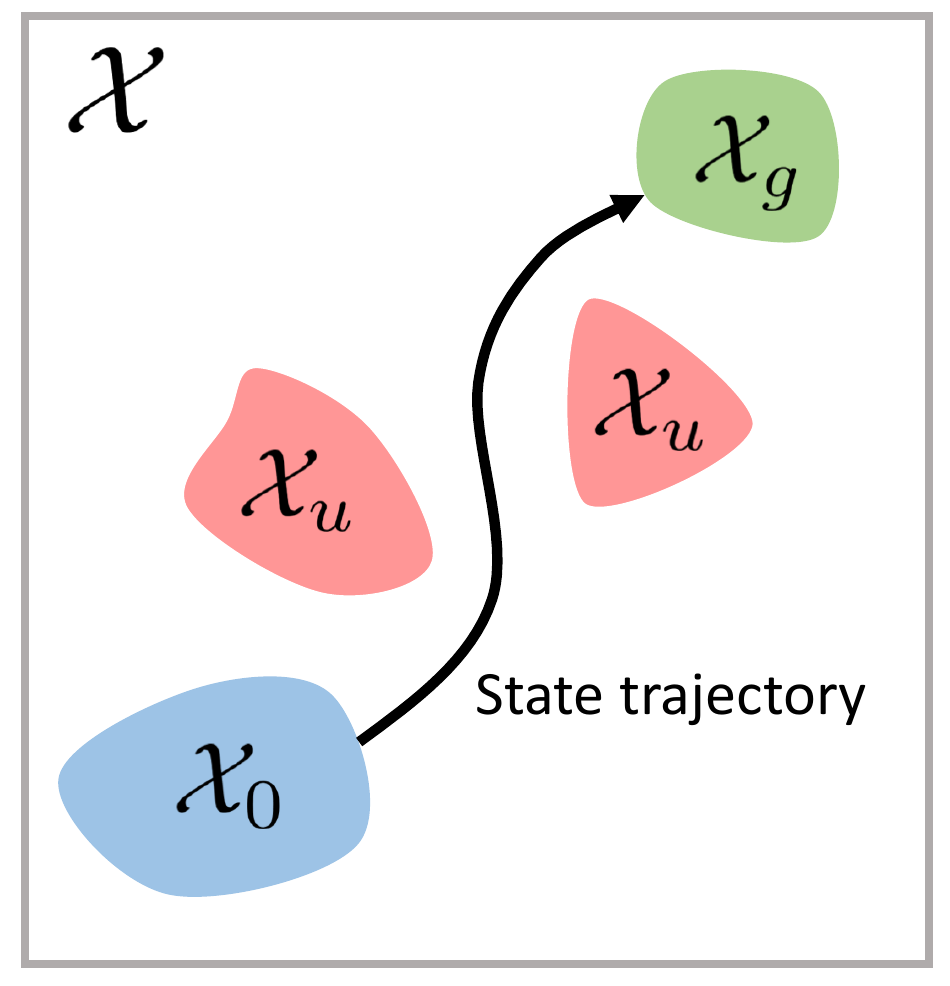}}
	\caption{Illustration of a safety and goal-reaching policy. $\mathcal{X}_0$, $\mathcal{X}_u$, and $\mathcal{X}_g$ are the initial, unsafe, and goal sets, respectively.}
	\label{fig.intro}
\end{wrapfigure}

In this paper, we distinguish  safety from stability: stability requires a policy to enable the system to have attractive or bounded behavior, while safety concerns that a policy must not drive the state of a system into any unsafe region. With this difference, we focus on learning a policy that is both safe and stable. As illustrated in Fig. \ref{fig.intro},  first  we guarantee that under the control policy, the trajectory of the system from an initial state set $\mathcal{X}_0$ \emph{asymptotically converges} to a specified goal state set $\mathcal{X}_g$. We here refer this property to as `goal-reaching', which is a generalization of stability. Second, the policy guarantees that the system, starting from \emph{any} state in an initial state set $\mathcal{X}_0$, must not enter into any unsafe state set $\mathcal{X}_{u}$, which we refer to as safety. This paper develops an approach to learn a policy of both safety and goal-reaching. The key here is to jointly learn the policy together with two certificate functions, where we use a neural barrier function \cite{wieland2007constructive} to account for the safety and a neural Lyapunov-like function to guarantee attractiveness of the system to a given goal set.

\subsection{Background and Related Work}

\textbf{Lyapunov stability.} \quad The Lyapunov theory  \cite{lyapunov1892general} is a systematic methodology to characterize the stability of a dynamical system. Given a valid Lyapunov function, the motion of a dynamical system is guaranteed to be bounded (or attractive to, depending on specific conditions) around an equilibrium point. However, finding a valid Lyapunov function is generally challenging. Although some methods, e.g., sum-of-square   \cite{prajna2002introducing} and learning from demonstrations \cite{ravanbakhsh2019learning}, attempt to propose a general solution, but they are limited to simple systems, such as requiring polynomial dynamics or restricting Lyapunov functions to be a weighted sum of features. In this paper, we develop a Lyapunov-like function, which can be viewed as a generalization of the Lyapunov theory, which provides a provable certificate for a controlled system to converge to   \emph{a state set} instead of an {equilibrium} point. 

\textbf{Lyapunov based safe learning.} \quad Due to the provable boundness  by the Lyapunov theory, a  set of recent work \cite{kolter2019learning,chang2019neural,richards2018lyapunov,chow2018lyapunov} treats  safe policy learning from the perspective of system stability. The core of these work is to jointly learn the policy and the Lyapunov function, which ensures the behavior of the controlled system constrained within a region of attraction considered as a safe region. In most of cases, these methods will lead to a policy which is not only safe but also stable in a sense of driving the system converging to a fixed (equilibrium) point. Such  definition and treatment of safety is usually conservative, because a safe policy does not necessarily have to be stable.

\textbf{Conservativeness comparison between optimality, stability, and safety.} \quad We include the optimal policy resulting from reinforcement learning and compare the conservativeness between the concepts of optimality, stability, and safety of a policy. First, optimality of a policy generally implies stability, and this is because the value function for an optimal policy can be directly viewed as a valid Lyapunov function since the Bellman equation satisfies all  conditions required for a valid Lyapunov function (note that the opposite claim does not necessarily hold) \cite{primbs1999nonlinear}. Second, definition of Lyapunov stability states the boundness of system behavior at an equilibrium point \cite{lyapunov1892general}, which can be understood as a more restrictive  safety. Thus, we can conclude that in terms of safety guarantee,  optimality is more conservative than stability, and  stability can be thought of as a more restrictive treatment of safety.

\textbf{Barrier functions for safe control.} \quad  Barrier functions \cite{prajna2004safety,wieland2007constructive,ames2016control,ames2019control} were introduced to the control field to account for safety requirement in control design. By providing a barrier function and defining safe/unsafe sets as level-sets of the barrier function,  trajectories of the system can be guaranteed to never cross the boundary of level sets.  In the learning field,  barrier functions are recently employed  to guarantee safety in learning of a policy or dynamics, such as \cite{taylor2019learning,wang2018safe,cheng2019end}. All these methods require a valid barrier function to be available.

\textbf{Learning barrier functions.} \quad Whenever the unsafe/safe sets are not given in the algebraic form using a barrier function, one cannot leverage the safety guarantee that  barrier functions provide. A very recent set of work has explored to learn barrier functions from data.  In  \cite{saveriano2019learning}, barrier functions are parameterized by \emph{linear functions}, and a method is developed to update the linear barrier functions incrementally. In \cite{srinivasan2020synthesis}, the authors  use the support vector machine to learn a barrier function in a supervised way. These  approaches however are restricted to the systems whose dynamics is \emph{affine} in control inputs. In those cases,  learning a barrier function is decoupled from finding a policy, and the learning process is reduced to solving a binary classification problem. Since the  policy is not considered in the process of  learning barrier function, the obtained barrier function may be no longer valid when the policy is included. Furthermore, the policy in these methods is usually solved using quadratic programming after the barrier function is learned;  even though safety is guaranteed, stability of the policy might not be satisfied.

\textbf{Our contributions.} \quad  First, instead of decoupling the process of learning a barrier function and obtaining a safe policy,  we jointly learn them from data to guarantee that the barrier function is valid and the policy is safe; and thus our approach is able to apply to more general non-linear systems (such as non-affine systems). Second, we develop a general Lyapunov-like function and  incorporate learning of such Lyapunov-like function into the policy learning process. As a result, the learned policy is guaranteed to be \emph{not only safe but also goal-reaching}  in a sense of driving the system to converge to a goal set. Here, the control policy,  barrier function, and  Lyapunov-like function are all represented by neural networks. Our results show that by jointly including these certificate functions (i.e., barrier and Lyapunov-like functions) in learning, we can finally obtain a policy that is both safe and goal-reaching.

\section{Problem Statement}

\textbf{Notations.}\quad
The set of non-negative real numbers are denoted by $\mathbb{R}^+_0$.
The gradient of a differentiable scalar function $V(\bm{x})$ is denoted as $\nabla V(\boldsymbol{x})$, i.e., $\nabla V(\boldsymbol{x})={\partial V}/{\partial \bm{x}}$. The distance from $\boldsymbol{x}$ to a set $\mathcal{S}$ is defined by $\norm{\boldsymbol{x}}_{\mathcal{S}}=\inf_{\bm{s}\in\mathcal{S}}\norm{\boldsymbol{x}{-}\bm{s}}$ ($\norm{\cdot}$ is the  Euclidean norm).
A continuous function $\alpha:[0,a)\rightarrow[0,+\infty)$ for some $a>0$ is said to belong to class $\mathcal{K}$ if it is strictly increasing and $\alpha(0)=0$. A continuous function  $\beta:(-b,a)\rightarrow(-\infty,+\infty)$ for some $a,b>0$ is said to belong to extended-class $\mathcal{K}$ if it is strictly increasing and $\beta(0)=0$. A continuous function $\gamma:[0,c)\times[0,\infty)\rightarrow[0,+\infty)$ for some $c>0$ is said to belong to class $\mathcal{KL}$, if for each fixed $s$, the mapping $\gamma(r,s)$ belongs to class $\mathcal{K}$ with respect to $r$ and for each fixed $r$, the mapping  $\gamma(r,s)$ is decreasing with respect to $s$  and $\gamma(r,s)\rightarrow 0$ as $s\rightarrow \infty$.

We consider the a  dynamical system:
\begin{equation}\label{dyn}
\dot{\boldsymbol{x}}=\boldsymbol{f}(\boldsymbol{x},\bm{u}), \quad\quad \boldsymbol{x}(0)=\bm{x}_0,
\end{equation}
where $\boldsymbol{x}\in \mathcal{X}\subseteq \mathbb{R}^n$ is the system state with  $\mathcal{X}$  defining the system  state space; $\boldsymbol{u}\in \mathcal{U}\subseteq\mathbb{R}^{m}$ is the control input  with $\mathcal{U}$  defining the  control space;  and the vector function $\bm{f}:\mathcal{X}\times\mathcal{U}\rightarrow\mathcal{X}$ is assumed to be continuously differentiable. For  (\ref{dyn}), we define a continuously differentiable policy
\begin{equation}\label{controller}
\bm{u}=\bm{u}(\bm{x}),
\end{equation}
which defines a  map from state to control, $\bm{u}:\mathcal{X}\rightarrow \mathcal{U}$. Equipped with the policy (\ref{controller}), the system (\ref{dyn}) becomes  a \emph{controlled (autonomous) system} $\dot{\bm{x}}=\bm{f}\big(\bm{x},{\bm{u}}({\bm{x}}))$, and we below  write as $\dot{\bm{x}}=\bm{f}_{\bm{u}}\big(\bm{x})$ for simplicity. Given an initial state $\bm{x}_0$, the motion of the controlled system $\dot{\bm{x}}=\bm{f}_{\bm{u}}\big(\bm{x})$ is a  trajectory denoted by $\bm{x}_{\boldsymbol{u}}(t)$ for $t \in \mathbb{R}^+_0$, which is a state of time with $\bm{x}_{\boldsymbol{u}}(0)=\boldsymbol{x}_0$.

The goal  in this paper is to find a policy (\ref{controller}) such that  the controlled system, starting from  $\bm{x}_0$, will be guaranteed to  (i)  \emph{reach a given goal state} and (ii) \emph{ensure safety}. We specifically illustrate these two requirements using Fig. \ref{fig.intro}. Define a set of initial states $\mathcal{X}_0\subseteq\mathcal{X}$, a set of unsafe states $\mathcal{X}_u\subseteq\mathcal{X}$ with $\mathcal{X}_u\cap\mathcal{X}_0=\emptyset$, and a set of goal states $\mathcal{X}_g\subseteq\mathcal{X}$ with $\mathcal{X}_u\cap\mathcal{X}_g=\emptyset$. Our goal is to find a policy $\bm{u}=\bm{u}(\bm{x})$ such that from any $\bm{x}_0\in\mathcal{X}_0$, the system's   trajectory $\bm{x}_{\bm{u}}(t)$ with  $t\in \mathbb{R}^{+}_0$ will finally reach the goal set $\mathcal{X}_g$  while avoiding entering into the unsafe set $\mathcal{X}_u$. More formally, we give the following definitions for the {safety and goal reaching} of a policy, respectively.

\begin{definition}[Safety of a policy]\label{defsafety}
	For the  system (\ref{dyn}) with a set of initial states $\mathcal{X}_0\subseteq\mathcal{X}$ and   a set of unsafe states $\mathcal{X}_u\subseteq\mathcal{X}$, we say a policy $\bm{u}=\bm{u}(\bm{x})$ is safe with respect to the unsafe set $\mathcal{X}_u$, if  there does NOT exist a trajectory $\bm{x}_{\bm{u}}(t)$ such that  $\bm{x}(0)\in\mathcal{X}_0$ and  $\bm{x}{(T)}\in\mathcal{X}_u$ for some $T\in \mathbb{R}^+_{0}$.
\end{definition}

\begin{definition}[Goal reaching of a policy]\label{defsatble}
	For the  system (\ref{dyn}) with a set of initial states $\mathcal{X}_0\subseteq\mathcal{X}$ and  a set of goal states $\mathcal{X}_g\subseteq\mathcal{X}$, we say a  policy $\bm{u}=\bm{u}(\bm{x})$ is  goal-reaching  with respect to the goal set $\mathcal{X}_g$, if there exists a $\mathcal{KL}$-function $\gamma$ such that for any $\bm{x}(0)\in {\mathcal{X}_0}$, 
	\begin{equation}
	\norm{\bm{x}_{\bm{u}}(t)}_{\mathcal{X}_g}\leq\gamma(\norm{\bm{x}(0)}_{\mathcal{X}_g},t) \quad \text{for all}\quad  t\in \mathbb{R}^{+}_0
	\end{equation}
\end{definition}
The definition for safety in Definition \ref{defsafety} states that a safe policy is the one that must not lead the system state, from initial state set $\mathcal{X}_0$, to the unsafe set $\mathcal{X}_u$. The definition of goal-reaching in Definition \ref{defsatble} requires the policy always to drive the system  state towards getting closer to the goal set $\mathcal{X}_{g}$. Based on these two formal definitions, we present the problem of interest of this paper:

\textbf{The problem of interest:} Given a system (\ref{dyn}) with   an initial state set $\mathcal{X}_{0}\subseteq\mathcal{X}$, an unsafe set $\mathcal{X}_{u}\subseteq\mathcal{X}$, and a goal set $\mathcal{X}_{g}\subseteq\mathcal{X}$, we aim to find a policy (\ref{controller}) that is both safe and  goal-achieving. 

\section{Theoretical Results}
We analyze the problem of interest by decomposing it to safe policy design and goal-reaching policy design, each of which will be discussed below. For the safety policy design, we use barrier functions, which were introduced to control systems in \cite{prajna2004safety,wieland2007constructive,ames2016control}. For the goal-reaching policy design, we modify the Lyapunov function \cite{lyapunov1892general,sontag1983lyapunov,lin1996smooth} and  develop a more general Lyapunov-like function. We refer to both  barrier and  Lyapunov-like functions as \emph{certificate functions}.

First, we adapt the barrier function used in \citep{prajna2004safety} to be the following:

\begin{definition}[Barrier function \cite{prajna2004safety}]\label{defbarrierfun}
	Consider a controlled system $\dot{\bm{x}}=\bm{f}_{\bm{u}}(\bm{x})$ with policy $\bm{u}:\mathcal{X}\rightarrow\mathcal{U}$ and  $\bm{f}_{\bm{u}}:\mathcal{X}\rightarrow\mathcal{X}$ both continuously differentiable, a set of initial states $\mathcal{X}_0\subseteq\mathcal{X}$, and a set of unsafe states $\mathcal{X}_u\subseteq\mathcal{X}$, A continuously differentiable function $B(\bm{x}):\mathcal{X}\rightarrow\mathbb{R}$ is called a control barrier function, if there exists an extended class-$\mathcal{K}$ function $\beta$ such that 
	\begin{subequations}\label{cbf}
		\begin{align}
		B(\bm{x})&\leq 0  \qquad\qquad\qquad \forall \bm{x}\in\mathcal{X}_0,\label{cbf.1}\\
		B(\bm{x})&> 0  \qquad \qquad\qquad\forall \bm{x}\in\mathcal{X}_u,\label{cbf.2}\\
		\nabla B(\bm{x})\bm{f}_{\bm{u}}(\bm{x})&\leq - \beta( B(\bm{x})) \qquad \, \forall \bm{x}\in\mathcal{X}.\label{cbf.3}
		\end{align}
	\end{subequations}
\end{definition}

Compared to the  barrier function defined in \citep{prajna2004safety,wieland2007constructive} which requires $\nabla B(\bm{x})\bm{f}_{\bm{u}}(\bm{x})\leq 0$ for all $\bm{x}\in\mathcal{X}$, we relax it in  (\ref{cbf.3}). Recall that $\beta(\cdot)$ is an extended class-$\mathcal{K}$ function, which could take negative values for $B(\bm{x})\leq 0$. Thus, we allow $\nabla B(\bm{x})\bm{f}_{\bm{u}}(\bm{x})>0$ for $B(\bm{x})<0$, which makes  (\ref{cbf.3}) more likely be fulfilled. The interpretation of the above barrier function is straightforward:  (\ref{cbf.1}) and (\ref{cbf.2})  state that  the unsafe and initial sets can be separated by the barrier function; and (\ref{cbf.3}) guarantees that  trajectory of the controlled system starting from $\mathcal{X}_0$ can only stay within the set $\{\bm{x}:B(\bm{x})\leq 0\}$ and cannot go out and enter into the unsafe set. The rigorous assertion for this is stated in as follows.

\begin{lemma}[Safe control theorem \cite{ames2016control}]\label{theoremsafecontrol}
	Given the controlled system $\dot{\bm{x}}=\bm{f}_{\bm{u}}(\bm{x})$ with  policy $\bm{u}:\mathcal{X}\rightarrow\mathcal{U}$ and  $\bm{f}_{\bm{u}}:\mathcal{X}\rightarrow\mathcal{X}$ both continuously differentiable, a set of initial states $\mathcal{X}_0\subseteq\mathcal{X}$, and a set of unsafe states $\mathcal{X}_u\subseteq \mathcal{X}$, if $B(\bm{x})$ is a  barrier function as in Definition \ref{defbarrierfun}, then the  policy $\bm{u}=\bm{u}(\bm{x})$ is safe with respect to $\mathcal{X}_u$, as defined in Definition \ref{defsafety}.
\end{lemma}

A proof of the above lemma can be found in  \cite{ames2016control} and also in Appendix \textcolor{red}{A}. We here give an intuition  behind Theorem \ref{theoremsafecontrol}. If these exists a barrier function $B(\bm{x})$, then any state $\bm{x}_{\bm{u}}(t)$ along the  trajectory  of the controlled system, starting from $\mathcal{X}_0$, will have $B(\boldsymbol{x}_{\boldsymbol{u}}(t))\leq0$ for all $t\in \mathbb{R}^+_0$ and $B(\boldsymbol{x}_{\boldsymbol{u}}(t))$ cannot become positive because the decreasing property (\ref{cbf.3}). This means that the state of the system will stay inside the set $\{\bm{x}:B(\bm{x})\leq0 \}$ cannot go out of it, thus of course is safe with respect to $\mathcal{X}_{u}$.

We next focus on find a certificate for the goal-reaching requirement in Definition \ref{defsatble}. To this end, we define the following Lyapunov-like function.

\begin{definition}[Lyapunov-like function]\label{deflyapunov}
	Consider a controlled system $\dot{\bm{x}}=\bm{f}_{\bm{u}}(\bm{x})$ with policy $\bm{u}:\mathcal{X}\rightarrow\mathcal{U}$ and  $\bm{f}_{\bm{u}}:\mathcal{X}\rightarrow\mathcal{X}$ both continuously differentiable, a set of initial states $\mathcal{X}_0\subseteq\mathcal{X}$, and a set of goal states $\mathcal{X}_g\subseteq \mathcal{X}$.
	A continuous differentiable function $V(\bm{x}):\mathcal{X}\rightarrow\mathbb{R}$ is said to be a Lyapunov-like function if  
	\begin{subequations}\label{equlyapu}
		\begin{align}
		&\emptyset\neq\{\bm{x}:V(\bm{x})\leq 0\}\quad \text{and}\quad\{\bm{x}:V(\bm{x})\leq 0\} \subseteq \mathcal{X}_g, \label{equlyapu.1}\\
		&	\nabla V(\bm{x})\bm{f}_{\bm{u}}(\bm{x})\leq - \beta( V(\bm{x})) \qquad \, \forall \bm{x}\in\mathcal{X},\label{equlyapu.2}
		\end{align}
	\end{subequations}
	for some extended class $\mathcal{K}$ function $\beta$.
\end{definition}

Note that the above Lyapunov-like function is more general than the classic Lyapunov function used in \cite{lyapunov1892general,chang2019neural,richards2018lyapunov,chow2018lyapunov}, where it requires $V(\bm{x})>0$ for $\bm{x}\in\mathcal{X}\backslash{0}$, $V(\bm{0})=0$, and $\nabla V(\bm{x})\bm{f}_{\bm{u}}(\bm{x})<0$. First, the above Lyapunov-like function allows for specifying a goal state set in addition to a fixed point. Second, the Lyapunov-like function do not necessarily require that $\nabla V(\bm{x})\bm{f}_{\bm{u}}(\bm{x})$ has to be always negative-definite, that is, $\nabla V(\bm{x})\bm{f}_{\bm{u}}(\bm{x})>0$ can happen for $\boldsymbol{x}\in \{\bm{x}:V(\bm{x})<0\}$; this will make the Lyapunov function less restrictive. Based on Definition \ref{deflyapunov}, we provide the following theorem, which states a certificate to guarantee the control policy is goal-reaching. 

\begin{theorem}[Goal-reaching control theorem]\label{theorm2}
	Given the controlled system $\dot{\bm{x}}=\bm{f}_{\bm{u}}(\bm{x})$ with  policy $\bm{u}:\mathcal{X}\rightarrow\mathcal{U}$ and $\bm{f}_{\bm{u}}:\mathcal{X}\rightarrow\mathcal{X}$  continuously differentiable, a set of initial states $\mathcal{X}_0\subseteq\mathcal{X}$, and a set of goal states $\mathcal{X}_g\subseteq \mathcal{X}$, if $L(\bm{x})$ is a barrier function as in Definition \ref{deflyapunov}, then the policy $\bm{u}=\bm{u}(\bm{x})$ is goal-reaching with respect to $\mathcal{X}_g$, as defined in Definition \ref{defsatble}.
\end{theorem}

A proof of the above theorem is given in Appendix \textcolor{red}{A}.
The assertion, however, is quite intuitive:  the existence of  $V(\bm{x})$ guarantees that the state along the system trajectory $\bm{x}_{\bm{u}}(t)$ is decreasing the value of $V(\bm{x})$ (see (\ref{equlyapu.2})), which implies that it is approaching  $\mathcal{X}_g$, thus goal-reaching by Definition \ref{defsatble}. 

Combining Theorems \ref{theoremsafecontrol} and \ref{theorm2}, we immediately obtain the following assertion stating that existence of barrier and Lyapunov-like certificates guarantees the policy is both safe and goal-achieving.

\begin{theorem}[Safe and goal-reaching control theorem]\label{keytheorem}
	Consider a controlled system $\dot{\bm{x}}=\bm{f}_{\bm{u}}(\bm{x})$ with policy $\bm{u}:\mathcal{X}\rightarrow\mathcal{U}$ and $\bm{f}_{\bm{u}}:\mathcal{X}\rightarrow\mathcal{X}$  both continuously differentiable. Given a set of initial states $\mathcal{X}_0\subseteq\mathcal{X}$, a set of unsafe states $\mathcal{X}_u\subseteq\mathcal{X}$,  and a set of goal states $\mathcal{X}_g\subseteq \mathcal{X}$, if there exist   a barrier function $B(\bm{x})$ as in Definition \ref{defbarrierfun} and a  Lyapunov-like function $V(\bm{x})$ as in Definition \ref{deflyapunov},  then the policy $\bm{u}=\bm{u}(\bm{x})$ is both safe with respect to $\mathcal{X}_u$ and goal-reaching with respect to $\mathcal{X}_g$. 
\end{theorem}

\section{Learning Neural Certificates for Safe and Goal-reaching Policies}

As stated in Theorem \ref{keytheorem},   obtaining the certificate functions, i.e., the barrier function $B(\bm{x})$ in Definition \ref{defbarrierfun} and  Lyapunov-like function $V(\bm{x})$ in Definition \ref{deflyapunov}, is the key step towards achieving a safe and goal-achieving policy. In this section, we  resort to  learning techniques and (deep) neural networks to find the policy together with learning the certificate functions. Note that to keep notation simply,  below we write $\bm{u}$ in  $\dot{\bm{x}}=\bm{f}_{\bm{u}}(\bm{x})$ as the parameter of a neural policy $\bm{u}=\bm{u}(\bm{x})$. 

\subsection{Construction of Neural Certificate Functions}

\textbf{Neural Barrier Functions.}\quad
We construct  the unknown barrier function as a multi-layer neural networks denoted as $B_{\bm{\theta}}(\bm{x})$, where the input is the system state and the output is a scalar value;  $\bm{\theta}$  denotes the parameter  of the  neural barrier function network; and we use $\tanh$ activation function in all layers to ensure the barrier function is differentiable with respect to its inputs. In order to successfully find a valid barrier function that satisfies the three conditions (\ref{cbf}) in  Definition \ref{defbarrierfun},  we propose to minimize the  following \emph{barrier certificate loss}
\begin{equation}\label{lossB}
\begin{aligned}
\text{Loss}_B(\boldsymbol{\theta},\boldsymbol{u})=
&\relu\Big(\sup_{\bm{x}\in\mathcal{X}_0}B_{\bm\theta}(\bm{x})\Big)+\relu\Big(\sup_{\bm{x}\in\mathcal{X}_u}(-B_{\bm\theta}(\bm{x}))+\epsilon\Big) \\
&+
\relu\Big(\sup_{\bm{x}\in\mathcal{X}}\nabla B_{\bm\theta}(\bm{x})\bm{f}_{\bm{u}}(\bm{x})+B_{\bm\theta}(\bm{x})\Big),
\end{aligned}
\end{equation}
where $\bm{u}$ is the parameter for a neural policy. Here the first term  on RHS of (\ref{lossB}) is used to penalize the violation of (\ref{cbf.1})  in the barrier definition; the second term  is to penalize the violation of (\ref{cbf.2}), where we use  a small pre-defined $\epsilon>0$ to guarantee the strict inequality in (\ref{cbf.2}) holds; and the third term is to penalize the violation of (\ref{cbf.3}), for which we  consider the extended class-$\mathcal{K}$ function $\beta(\cdot)$ is $x=\beta(x)$. It is clear that a minimal zero-loss value  of the above barrier-certificate loss function will indicate a valid barrier function ${B}_{\boldsymbol{\theta}}(\boldsymbol{x})$, which satisfies all conditions (\ref{cbf}), and the learned policy $\bm{u}$ thus is  safe.

One difficulty of minimizing the above barrier-certificate loss is that it contains three inner optimization problems for the neural barrier network, which might be computationally intractable for large-dimensional or deep networks. In implementation, we thus use the following \emph{empirical barrier certificate risk} to approximate the above loss function:
\begin{equation}\label{riskB}
\begin{aligned}
l_{B}(\bm\theta,\bm{u})=&\relu\left(\frac{1}{N_0}\sum\nolimits_{i=1}^{N_0}
B_{\bm\theta}(\bm{x}_0^i)\right)+\relu\left(\frac{1}{N_u}\sum\nolimits_{i=1}^{N_u}
B_{\bm\theta}(\bm{x}_u^i)\right)\\
&+
\relu\left(\frac{1}{N}\sum\nolimits_{i=1}^{N}\big(\nabla B_{\bm\theta}(\bm{x}^i)\bm{f}_{\bm{u}}(\bm{x}^i)+B_{\bm\theta}(\bm{x}^i)\big)\right).
\end{aligned},
\end{equation}
where $\{\boldsymbol{x}_0^i\}_{i=1}^{N_0}$, $\{\boldsymbol{x}_u^i\}_{i=1}^{N_u}$, and $\{\boldsymbol{x}^i\}_{i=1}^{N}$ are the $N_0$ samples of states sampled from the initial set $\mathcal{X}_0$, $N_u$ samples of states from $\mathcal{X}_u$, and $N$ samples of states from the entire state domain $\mathcal{X}$, respectively.

The above relaxation, however, means that the minimal  barrier-certificate risk (\ref{riskB}) may not always guarantee the minimal (zero) loss of the barrier-certificate loss  (\ref{lossB}). Thus, we need to have a verification step for the obtained barrier function $B_{\boldsymbol{\theta}}(\bm{x})$ and policy $\boldsymbol{u}$, which will be discussed later.

\textbf{Neural Lyapunov-like Functions.} \quad We construct  the neural Lyapunov-like function $V_{\bm{\omega}}(\bm{x})$, where its input is the state and output is a scalar value; $\bm{\omega}$ is the  neural network parameter; and we use $\tanh$ activation  in all layers to ensure $V_{\bm{\omega}}(\bm{x})$ is differentiable with respect to its inputs. To satisfy the two conditions (\ref{equlyapu}) of the Lyapunov-like function, we propose the following \emph{Lyapunov-like certificate loss}:
\begin{equation}\label{lossV}
\begin{aligned}
\text{Loss}_V(\boldsymbol{\omega},\boldsymbol{u})=
&\relu\Big(\sup_{\bm{x}\in\mathcal{\bar{X}}_g}V_{\bm\omega}(\bm{x})\Big)+\relu\Big(\sup_{\bm{x}\in{\mathcal{X}\textbackslash\mathcal{\bar{X}}_g}}(-V_{\bm\omega}(\bm{x}))\Big) \\
&+
\relu\Big(\sup_{\bm{x}\in\mathcal{X}}\nabla V_{\bm\omega}(\bm{x})\bm{f}_{\bm{u}}(\bm{x})+V_{\bm\omega}(\bm{x})\Big),
\end{aligned}
\end{equation}
where $\mathcal{\bar{X}}_g$ is any non-empty subset of the goal set $\mathcal{X}_g$ (recall the set conditions in (\ref{equlyapu.1}) in Definition \ref{deflyapunov}), namely, $\mathcal{\bar{X}}_g\neq\emptyset$ and $\mathcal{\bar{X}}_g\subseteq\mathcal{X}_g$, which can be flexibly chosen. The first two terms on RHS of (\ref{lossV}) are to penalize the violation of condition (\ref{equlyapu.1}) in Definition \ref{deflyapunov}, and the third term is to penalize the violation of condition (\ref{equlyapu.2}),  where we  consider the extended class-$\mathcal{K}$ function $\beta(\cdot)$ is $x=\beta(x)$. It is clear that the minimal zero-loss value of (\ref{lossV}) indicates a valid Lyapunov-like function $V_{\bm{\omega}}(\bm{x})$ satisfying all conditions in (\ref{equlyapu}) and a goal-reaching policy $\bm{u}$.

In practice, due to the complexity of directly minimizing the above Lyapunov-like certificate loss (\ref{lossV}), we choose its simple relaxation described as below. First, we construct $V_{\bm{\omega}}(\bm{x})$ as
\begin{equation}\label{vfunction}
V_{\bm{\omega}}(\bm{x})=\boldsymbol{\phi}_{\bm\omega}(\bm{x})^\prime\boldsymbol{\phi}_{\bm\omega}(\bm{x}),
\end{equation} 
where $\boldsymbol{\phi}_{\bm\omega}(\bm{x}):\mathbb{R}^{n}\rightarrow\mathbb{R}^o$  can be a multiple neural layer. One property of this construction is that $V_{\bm{\omega}}(\bm{x})\geq 0$ holds  for all $\boldsymbol{x}\in\mathcal{X}$. Second, we define the \emph{empirical Lyapunov-like certificate risk}:
\begin{equation}\label{riskV}
\begin{aligned}
l_{V}(\bm\theta,\bm{u})=\relu\left(\frac{1}{N_g}\sum_{i=1}^{N_g}
V_{\bm\omega}(\bm{x}_g^i)\right)+
\relu\left(\frac{1}{N}\sum_{i=1}^{N}\nabla V_{\bm\omega}(\bm{x}^i)\bm{f}_{\bm{u}}(\bm{x}^i)+V_{\bm\omega}(\bm{x}^i)\right)
\end{aligned},
\end{equation}
where $\{\boldsymbol{x}_g^i\}_{i=1}^{N_g}$ and $\{\boldsymbol{x}^i\}_{i=1}^{N}$ are $N_g$ samples of states sampled from the chosen goal subset $\mathcal{\bar{X}}_g$, and $N$ samples of states from the entire domain $\mathcal{X}$, respectively. Corresponding to the original (\ref{lossV}), the RHS second term of (\ref{lossV}) is removed because now the parameterized Lyapunov-like function  $V_{\bm{\omega}}(\bm{x})$ in (\ref{vfunction}) is non-negative. Again, the minimum of the above Lyapunov-like certificate risk may not guarantee that the two conditions (\ref{equlyapu}) of the  Lyapunov-like function are strictly satisfied. Thus, we must verify the learned neural Lyapunov-like Function, as discussed in the next subsection.

\subsection{Joint Learning of Neural Policies and Neural Certificates}

As we have defined the barrier certificate risk $l_{B}(\bm\theta)$ in (\ref{riskB})  and the Lyapunov-like certificate risk $l_V(\bm{\omega})$ in (\ref{riskV}), we can now join them together to define the \emph{total certificate risk} below as our optimization objective:
\begin{equation}\label{loss}
\min_{\bm{\theta},\bm{\omega},\bm{u}} l(\bm{\omega},\bm\theta,\bm{u})=\min_{\bm{\theta},\bm{\omega},\bm{u}}\big(l_{B}(\bm\theta,\bm{u})+l_V(\bm{\omega}, \bm{u})\big),
\end{equation}
where $\bm{u}$ denotes the parameter of a neural network policy. Optimizing the above total certificate risk $l(\bm{\omega},\bm\theta,\bm{u})$ is a joint training of a policy neural network and two certificate neural networks (the  barrier neural network and Lyapunov-like neural network). We note that the total certificate risk $l(\bm{\omega},\bm\theta,\bm{u})$ is positive semi-definite, and any $({\bm{\theta},\bm{\omega},\bm{u}})$ corresponding to true certificate functions and policy satisfy $l({\bm{\theta},\bm{\omega},\bm{u}})=0$. Thus the true certificate functions defines a global minimizer for  $l(\bm{\omega},\bm\theta,\bm{u})$.

\textbf{Verification of Learned Neural Certificates.}\quad
Since  $l_{B}(\bm\theta,\bm{u})$ and $l_V(\bm{\omega}, \bm{u})$ are relaxations of the barrier certificate loss (\ref{lossB}) and Lyapunov-like certificate loss (\ref{lossV}), respectively, we cannot immediately say that the learned neural barrier and Lyapunov-like certificates, which are the minimizer of  (\ref{loss}), are valid in a sense of satisfying Definition \ref{defbarrierfun} and Definition \ref{deflyapunov}, respectively. Thus, we must perform a verification step. To do so, we can employ multiple techniques, such as SMT algorithms used in \cite{chang2019neural} and the Lipschitz method used in \cite{richards2018lyapunov}. In our case, we use a simple verification method similar to \cite{richards2018lyapunov}. Specifically, for example, to verify a scalar function $h(\bm{x})<0$ for $\bm{x}\in\mathcal{D}$, it is sufficient to check the tightened condition $
h(\boldsymbol{x})<-L_h\tau=-\epsilon_1$ at finite set of points that  can cover $\mathcal{D}$. Here $L_{h}>0$ is the Lipschiz constant of $h(\bm{x})$ and $\tau>0$ is a measure of how densely the points are located over  $\mathcal{D}$ \cite{berkenkamp2017safe}. In our experiments, we  use a  discretization of $\mathcal{D}$ (possible adaptive discretization will yield better scaling to higher-dimensional state space \cite{richards2018lyapunov}). To verify the first condition (\ref{equlyapu.1}) of the Lyapunov-like function, since $V_{\boldsymbol{\omega}}(\bm{x})$ is constructed to be non-negative in (\ref{vfunction}), we only verify if $ \min_{i=1,2,\cdots} V_{\boldsymbol{\omega}}(\bm{x}^i_{g})\leq \epsilon_2$, where $\epsilon_2>0$ is a small constant parameter that bounds the tolerable numerical error, and $\bm{x}^i_g$ are the points in the discretization of $\mathcal{\bar{X}}_g$. In our experiments, we set $\epsilon_1=\epsilon_2$.

\section{Empirical Experiments}

We show the effectiveness of the proposed method to learn safe and goal-reaching policy on various nonlinear systems, including  pendulum, cartpole, wheeled vehicle path following, and UAV systems. In all experiments, we set the learning rate $10^{-3}$ and use ${N}=500$ samples from the state space $\mathcal{X}$, ${N_u}=500$ samples from the unsafe state set $\mathcal{X}_u$, ${N_0}=500$ samples from the initial state set $\mathcal{X}_0$, and ${N_g}=500$ samples  from the goal state set $\mathcal{X}_g$. All samples are subject to uniform distributions. For validation of certificate functions, we discretize each set with $500$ discretization points, and set parameter $\epsilon_1=\epsilon_2=10^{-4}$. We observe that the choice of the parameter $\epsilon_1$ and $\epsilon_2$ in the range of $[10^{-1}, 10^{-5}]$ do not affect the finally learned control and certificate functions.  More experiment information is provided in Appendix \textcolor{red}{B}.

\begin{figure}
	\begin{subfigure}{.33\textwidth}
		\centering
		\includegraphics[width=\linewidth]{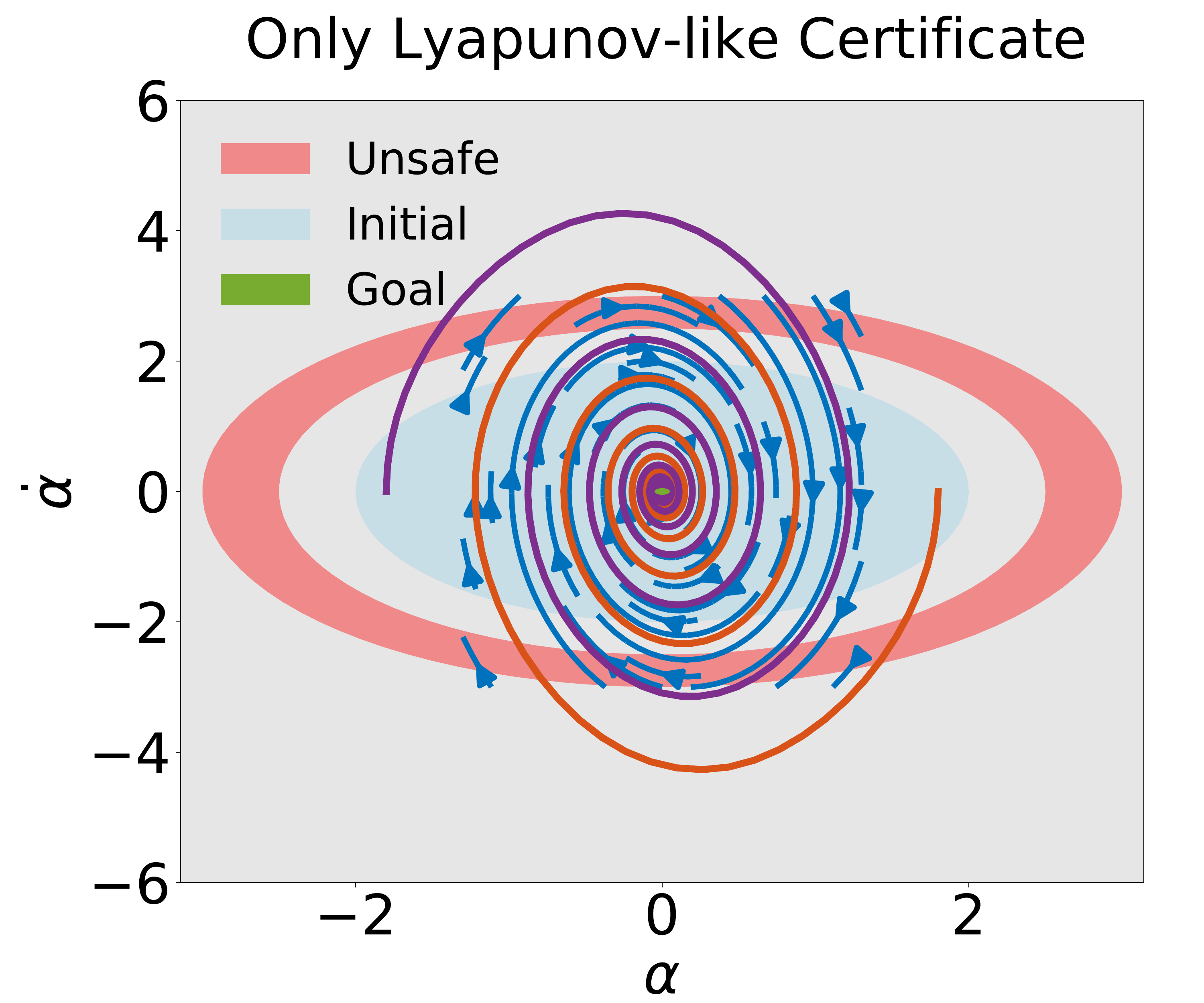}
		\caption{Only using Lyapunov-like}
		\label{pendulum.1}
	\end{subfigure}
	\begin{subfigure}{.33\textwidth}
		\centering
		\includegraphics[width=\linewidth]{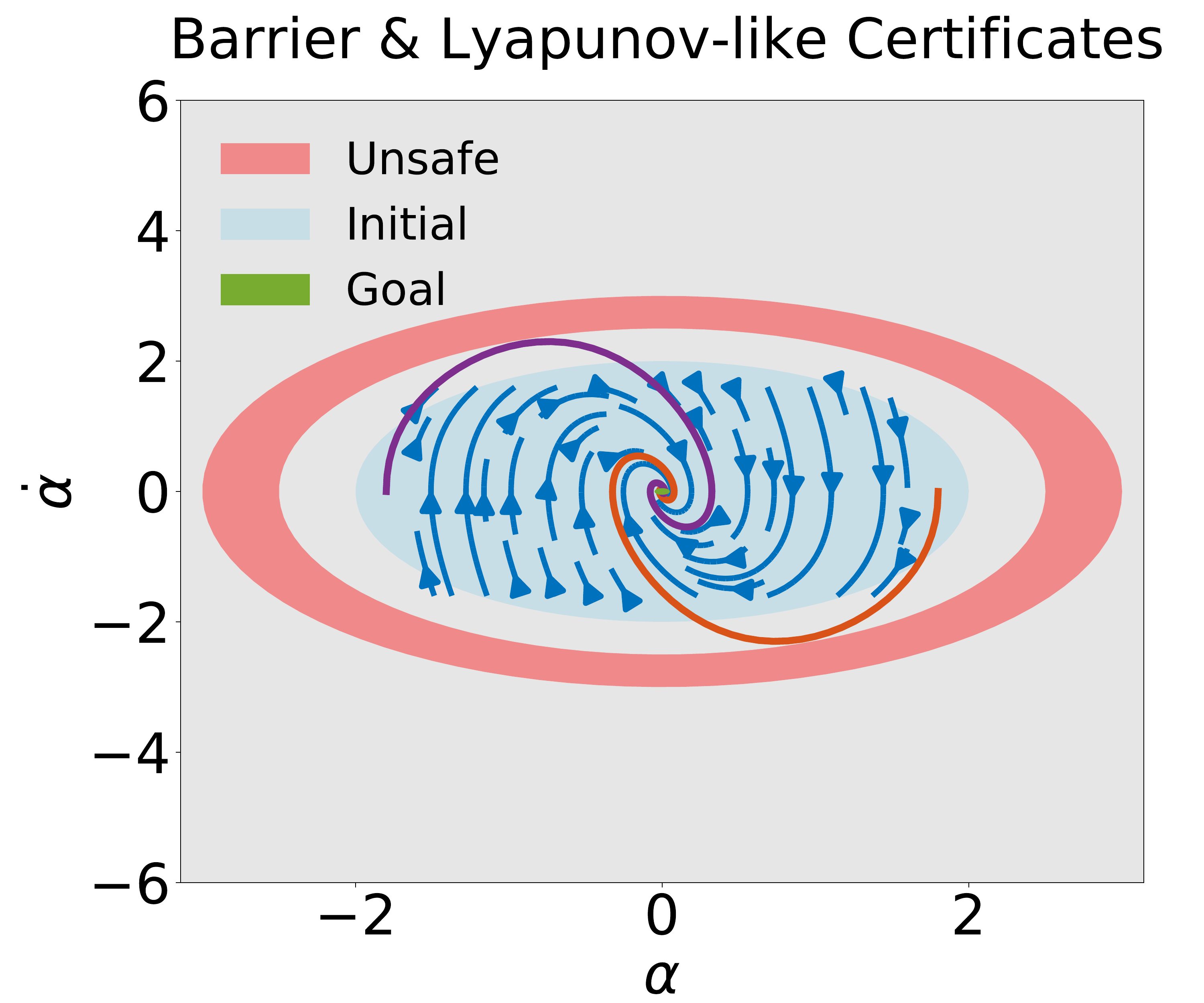}
		\caption{Barrier \& Lyapunov-like}
		\label{pendulum.2}
	\end{subfigure}
	\begin{subfigure}{.33\textwidth}
		\centering
		\includegraphics[width=\linewidth]{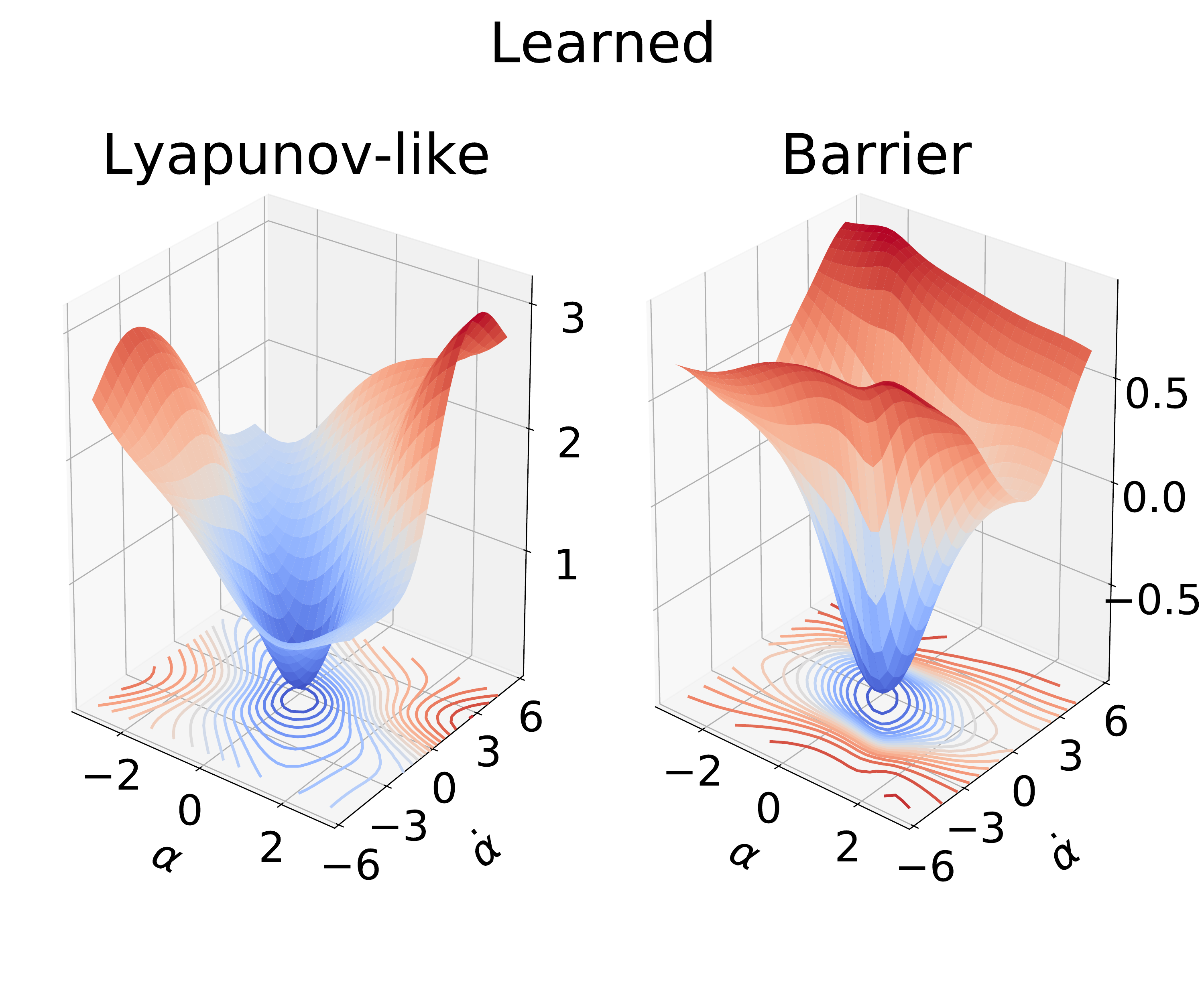}
		\caption{Learned neural certificates}
		\label{pendulum.3}
	\end{subfigure}
	\caption{Learning policies of a pendulum system: (a) simulation of trajectories of the system under the policy learned only using neural Lyapunov-like certificate function; (b) simulation of trajectories of the system under the  policy learned using both barrier and Lyapunov-like certificate functions, and (c) the learned Lyapunov-like function and barrier function for (b).} 
	\label{pendulum}
\end{figure}

\textbf{Pendulum system.}\quad We first test the  method using a pendulum system, where the state variable defined is $\bm{x}=[\alpha,\dot{\alpha}]^\prime$ with $\alpha$ being the angle between the  gravity and the pendulum, and the control variable ${u}$ is the torque applied to the pivot of the pendulum. Define the state space as $\mathcal{X}=\{\bm{x}:\bm{x}_{\text{lb}}\leq \bm{x}\leq \bm{x}_{\text{ub}}\}$ with $\bm{x}_{\text{lb}}=[-\pi, -5]^\prime $ and $\bm{x}_{\text{ub}}=[\pi, 5]^\prime $, the unsafe state set $\mathcal{X}_u=\{\bm{x}: 2.5\leq \norm{\bm{x}}_2\leq3 \}$, the goal state set  $\mathcal{X}_g=\{\bm{x}: \norm{\bm{x}}_2=0 \}$, and the initial state set  $\mathcal{X}_0=\{\bm{x}: \norm{\bm{x}}_2\leq 2 \}$. All these sets are indicated in Fig. \ref{pendulum.1} and \ref{pendulum.2}  using different colors.

We first  learn the policy together with  only neural Lyapunov-like function, i.e., by minimizing (\ref{riskV}), and the results are in Fig. \ref{pendulum.1}. From the results, we observe that with the  Lyapunov-like certificate, the obtained policy is guaranteed to be goal-reaching, as shown in Fig.  \ref{pendulum.1}, where the trajectory  converges to the goal set (here is origin). However, during the motion, the system trajectory has entered into the unsafe set (labeled by red region), and  this indicates that only using Lyapunov-like certificates, we cannot guarantee the safety of the policy. Next, we learn the policy with learning both the barrier and Lyapunov-like functions, and the results are in Fig. \ref{pendulum.2}, from which we  observe that the policy in this case drives the system towards the goal while successfully avoiding the unsafe set.  This is  because we have successfully learned the Lyapunov-like and barrier functions (shown in Fig. \ref{pendulum.3}), and they provide a provably  guarantee for the safe and stability of the policy. Comparing the results in Fig. \ref{pendulum.1} and \ref{pendulum.2}, we can conclude that  by jointly learning both the barrier and Lyapunov-like functions, we can achieve a policy not only safe but also goal-reaching (stable).

\begin{figure}[h]
	\begin{subfigure}{.33\textwidth}
		\centering
		\includegraphics[width=\linewidth]{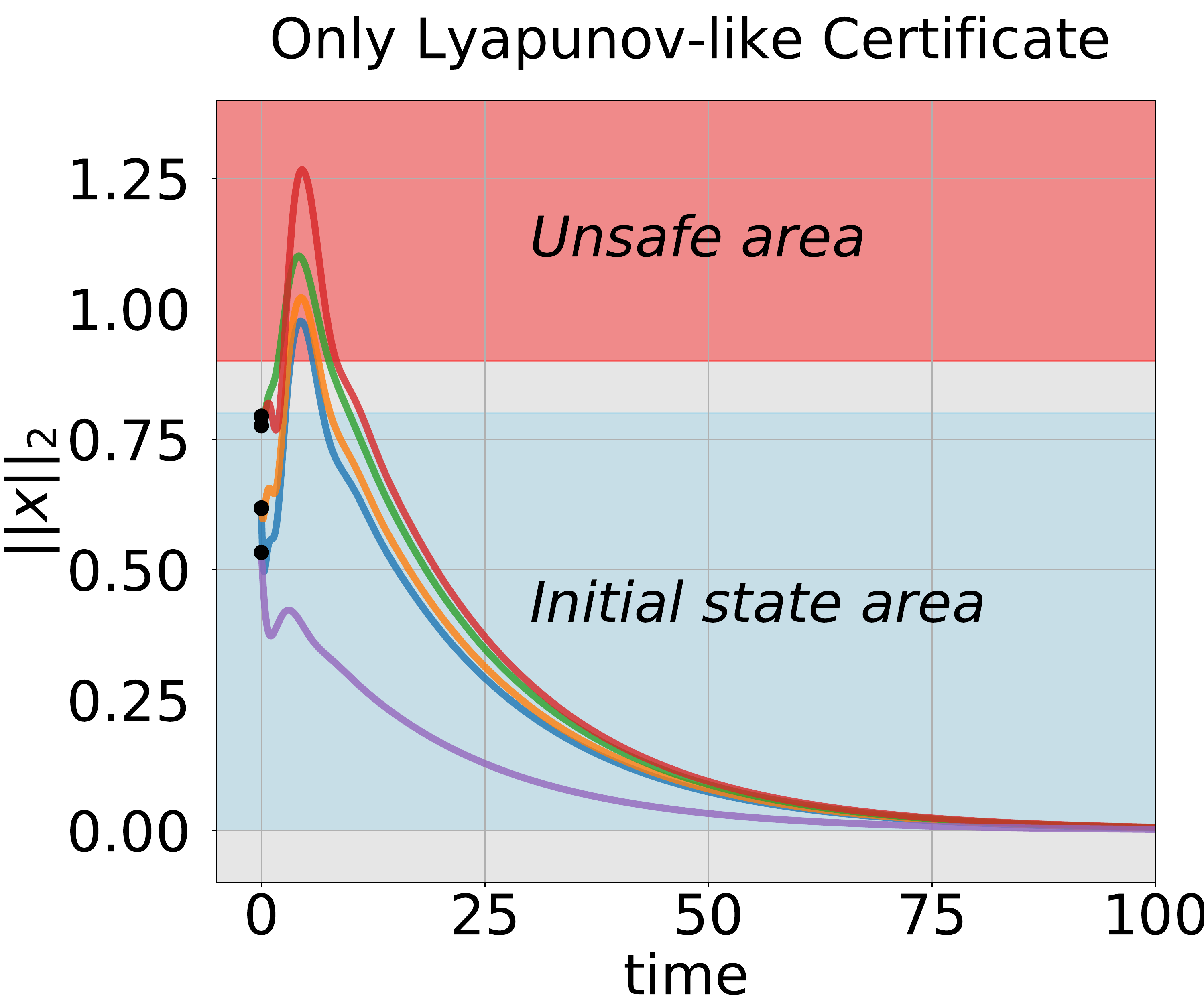}
		\caption{Only using Lyapunov-like}
		\label{cartpole.1}
	\end{subfigure}
	\begin{subfigure}{.33\textwidth}
		\centering
		\includegraphics[width=\linewidth]{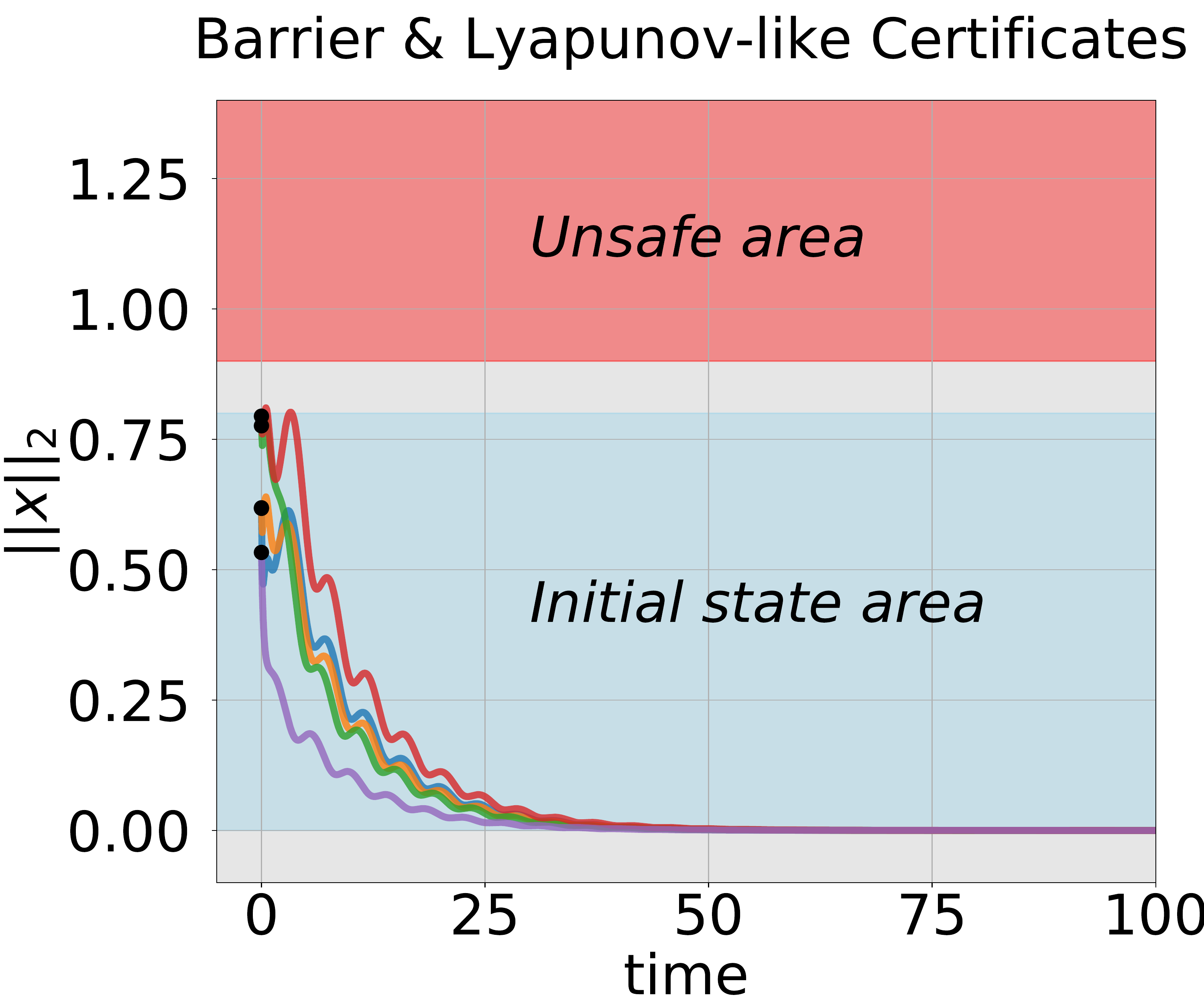}
		\caption{Barrier \& Lyapunov-like}
		\label{cartpole.2}
	\end{subfigure}
	\begin{subfigure}{.33\textwidth}
		\centering
		\includegraphics[width=\linewidth]{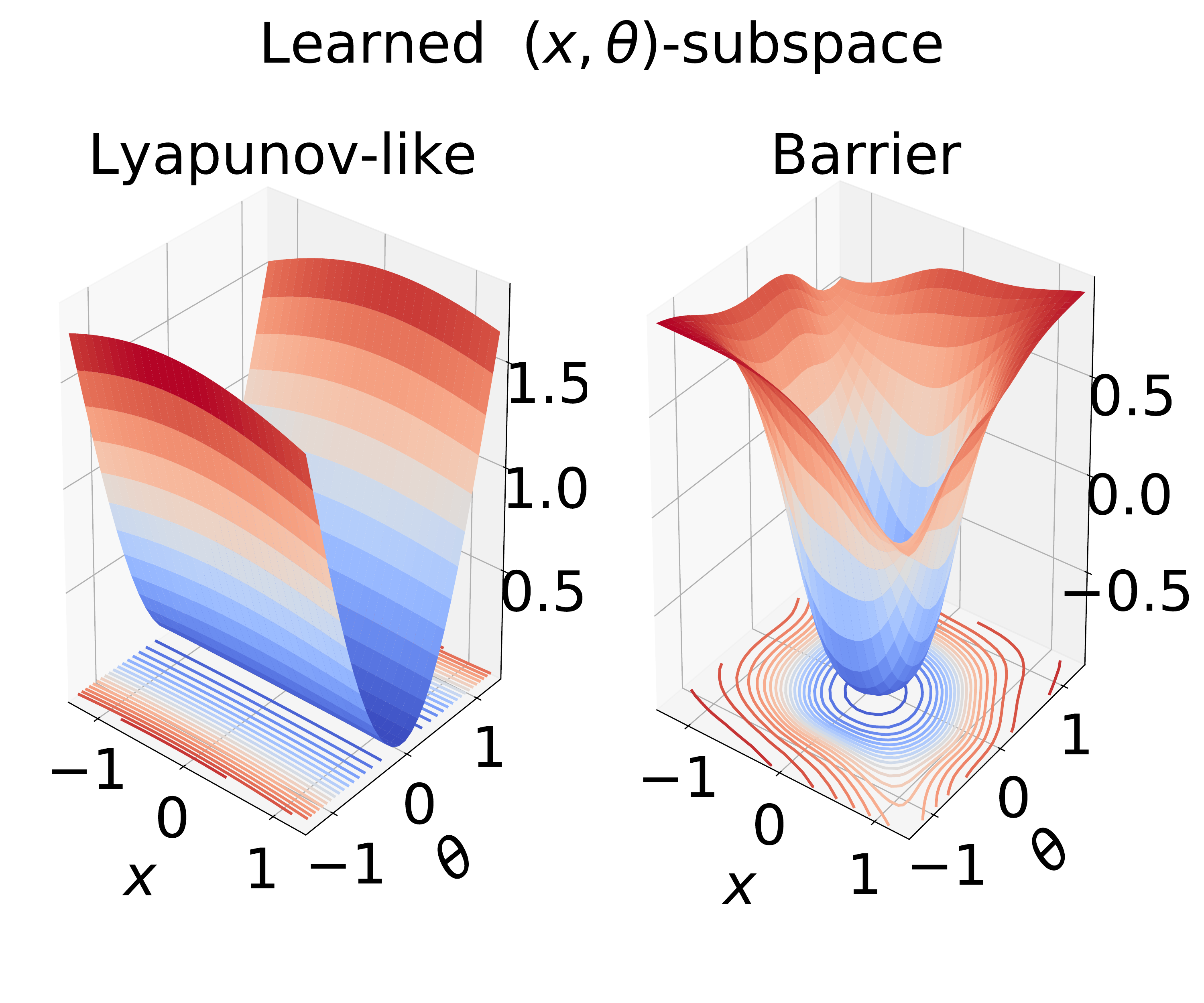}
		\caption{Learned neural certificates}
		\label{cartpole.3}
	\end{subfigure}
	\caption{Learning policies for cartpole system: (a) simulation of trajectories ($\norm{\bm{x}(t)}_2$ versus $t$)  of the system under the policy learned only with Lyapunov-like certificate; (b) simulation  of trajectories ($\norm{\bm{x}(t)}_2$ versus $t$) of the system under the  policy learned with both barrier and Lyapunov-like certificates; and (c)  the learned Lyapunov-like function and barrier function for (b) (note that we only show the 3D surface plots in the ($x,\theta$) subspace).} 
	\label{cartpole}
\end{figure}

\textbf{Cartpole system.} \quad We next test the proposed method to learn a safe policy to stabilize a cart-pole system. The system state is defined as $\boldsymbol{x}=[x,\theta,\dot{x},\dot{\theta}]^\prime$, where $x$ is the position of the cart and $\theta$ is the angle between pole and gravity.  Define the state domain $\mathcal{X}=\{\bm{x}:\bm{x}_{\text{lb}}\leq \bm{x}\leq \bm{x}_{\text{ub}}\}$ with $\bm{x}_{\text{lb}}=[\text{-}1.3, \text{-}1.3, \text{-}1.3, \text{-}1.3]^\prime $ and $\bm{x}_{\text{ub}}=[1.3, 1.3, 1.3, 1.3]^\prime $, the unsafe state set $\mathcal{X}_u=\{\bm{x}: 0.9\leq \norm{\bm{x}}_2\leq 1.3 \}$, the goal state set  $\mathcal{X}_g=\{\bm{x}:  \norm{\bm{x}}_2=0 \}$, and the initial state set  $\mathcal{X}_0=\{\bm{x}: \norm{\bm{x}}_2\leq 0.8 \}$. $\mathcal{X}_u$ and $\mathcal{X}_0$ are labeled in Fig. \ref{cartpole.1} and \ref{cartpole.2} using different colors.

First, we learn the policy with only the neural Lyapunov-like function. The simulation of the system under the learned policy is in Fig. \ref{cartpole.1}, where  we show five trajectories (labeled by different colors with black dots denoting initial states) of the system under the learned policy.  The results  show that all of the trajectories go towards to the goal set (i.e. converging to $\mathcal{X}_g=\{\bm{x}:  \norm{\bm{x}}_2=0 \}$), but four of them  have crossed into the unsafe area $\mathcal{X}_u=\{\bm{x}: 0.9\leq \norm{\bm{x}}_2\leq 1.3 \}$. This shows that  only using Lyapunov-like certificate can guarantee the goal-reaching (stability), but cannot ensure the safety. Next, we learn the policy with jointly learning both the Lyapunov-like and barrier functions, and the simulation results for the learned policy in this case are in Fig. \ref{cartpole.2} (the initial states are the same with \ref{cartpole.1}). The results show that with the learned Lyapunov-like and barrier  certificates, the obtained policy is both safe and goal-reaching. Fig. \ref{cartpole.3} shows the  learned neural Lyapunov-like and barrier functions, where we only plot the surface in the subspace of ($x,\theta$) (full state is four dimensional).

\begin{figure}[h]
	\begin{subfigure}{.25\textwidth}
		\centering
		\includegraphics[width=\linewidth]{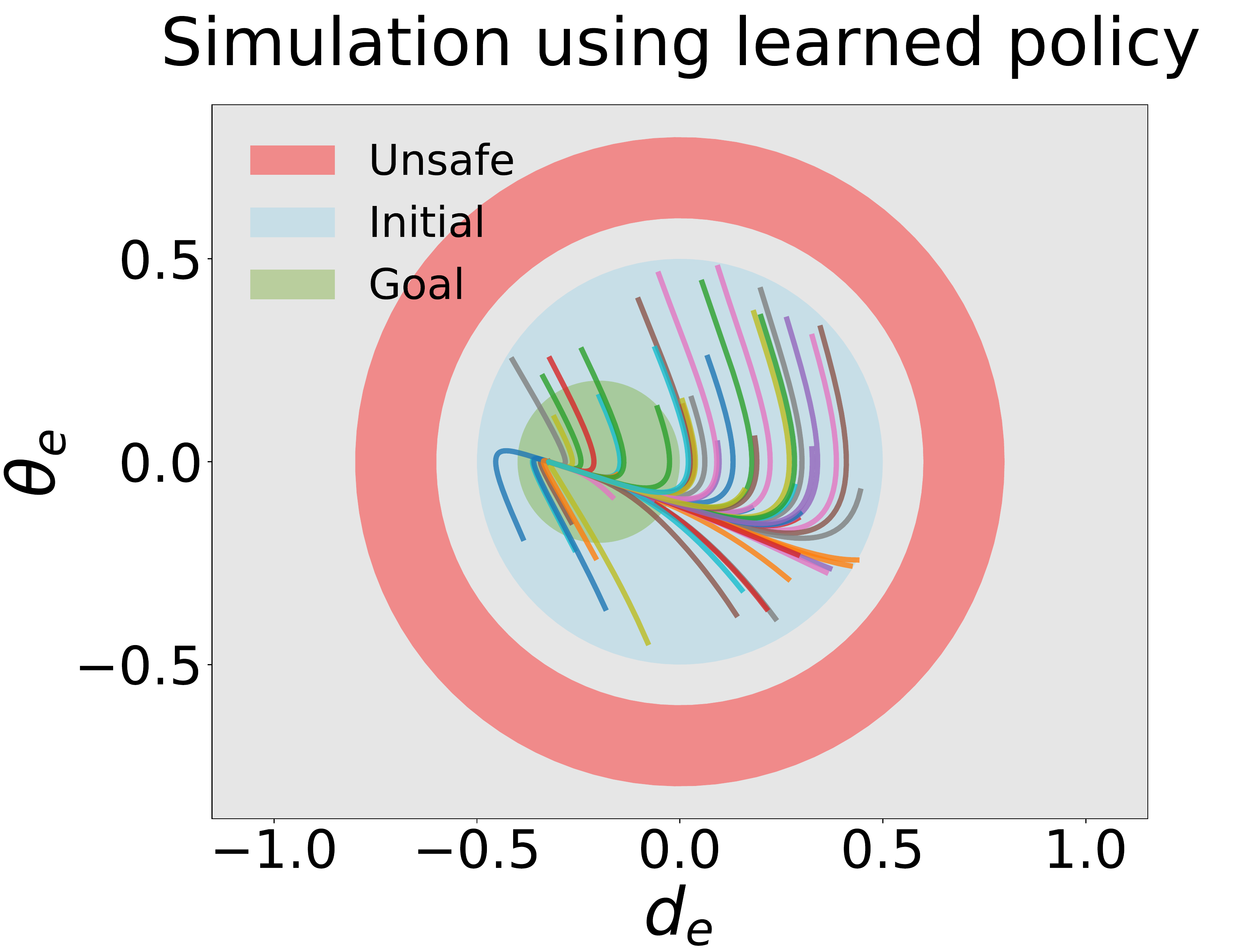}
		\caption{Vehicle simulation}
		\label{bicycle.1}
	\end{subfigure}%
	\begin{subfigure}{.25\textwidth}
		\centering
		\includegraphics[width=\linewidth]{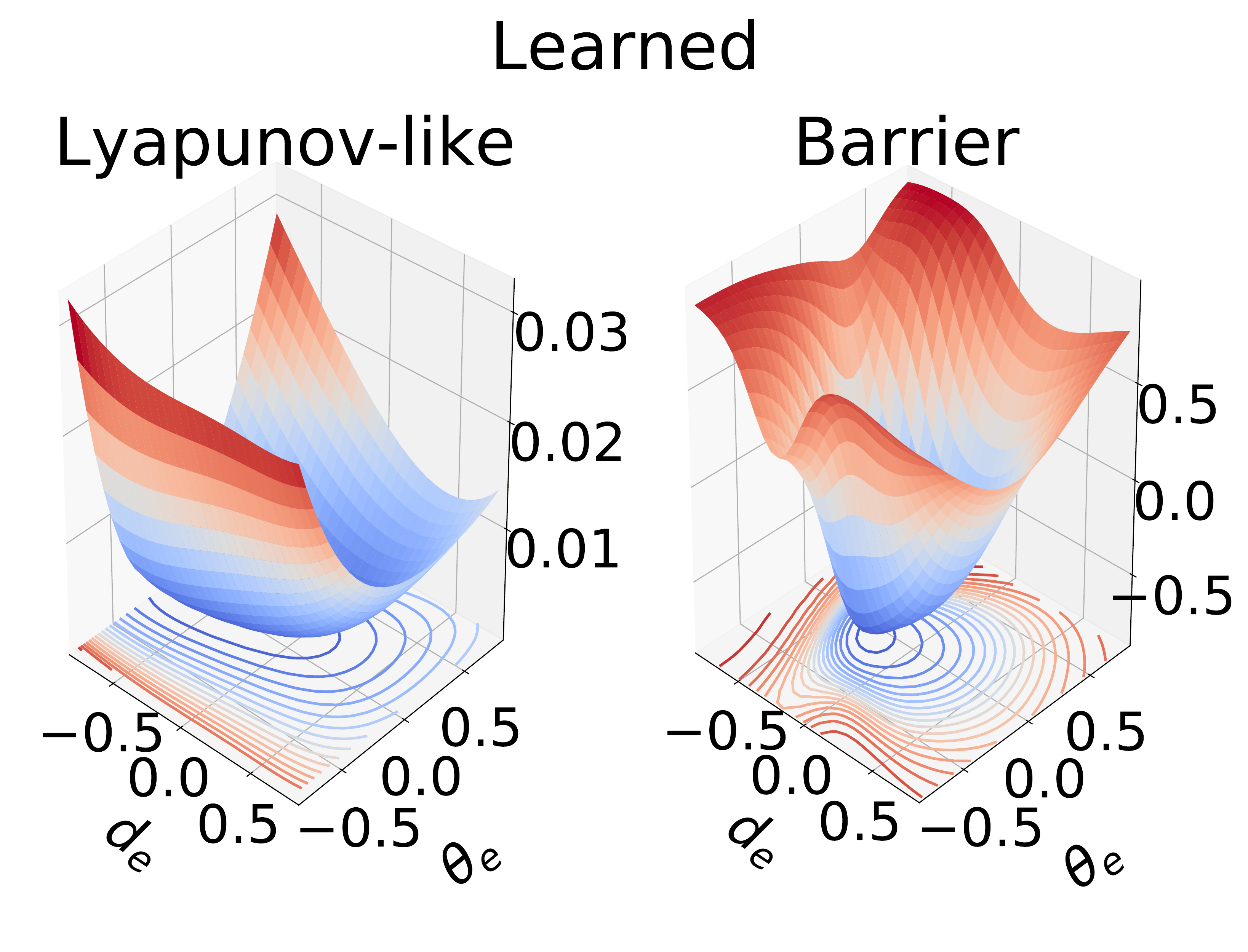}
		\caption{Neural certificates}
		\label{bicycle.2}
	\end{subfigure}%
	\begin{subfigure}{.25\textwidth}
		\centering
		\includegraphics[width=\linewidth]{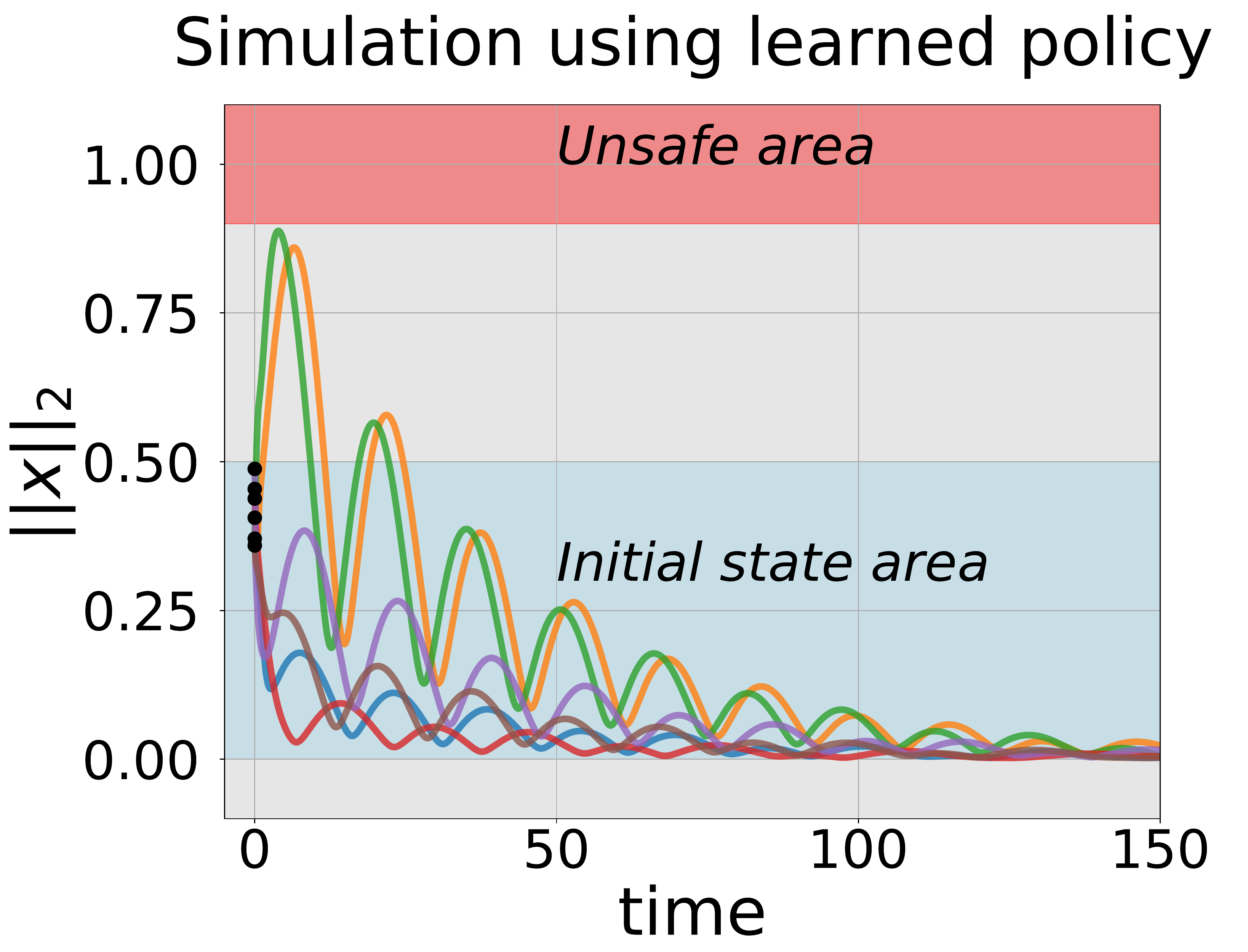}
		\caption{UAV simulation}
		\label{uav.1}
	\end{subfigure}%
	\begin{subfigure}{.25\textwidth}
		\centering
		\includegraphics[width=\linewidth]{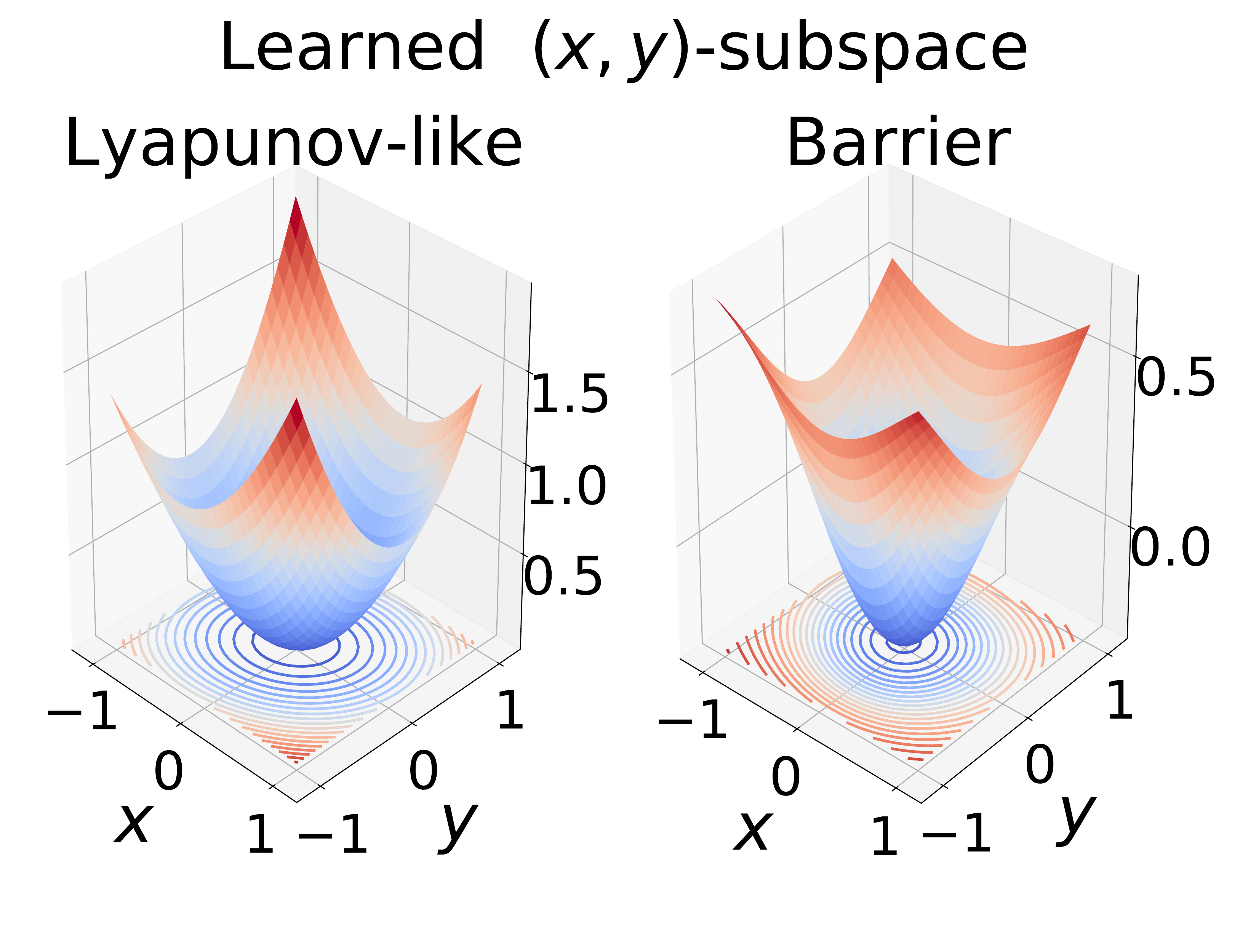}
		\caption{Neural certificates}
		\label{uav.2}
	\end{subfigure}
	\caption{(a) Simulation of trajectories of the vehicle path tracking system under the  policy learned with both neural Lyapunov-like and barrier certificates; (b) the learned neural Lyapunov-like function and barrier function for (a);  (c) simulation  of trajectories ($\norm{\bm{x}(t)}_2$ versus $t$) of the UAV under the policy learned using both barrier and Lyapunov-like certificates, and here the black dots denote the norm of different initial states $\norm{\bm{x}_0}$, and (d)  the learned Lyapunov-like function and barrier function for (c), here note that we have only show 3D surface plots in the ($x,y$) subspace.} 
	\label{bicycle-uav}
\end{figure}

\textbf{Vehicle path tracking.} \quad Next, we  test the method on a wheeled vehicle tracking a reference path \cite{chang2019neural}. We define the system state $\bm{x}=[d_e, \theta_e]^\prime$ with $d_e$ and $\theta_e$ being the distance and angle difference between the vehicle and a reference path, respectively.  Define the state space $\mathcal{X}=\{\bm{x}:\bm{x}_{\text{lb}}\leq \bm{x}\leq \bm{x}_{\text{ub}}\}$ with $\bm{x}_{\text{lb}}=[\text{-}0.8, \text{-}0.8]^\prime $ and $\bm{x}_{\text{ub}}=[0.8, 0.8]^\prime $, the  unsafe state set $\mathcal{X}_u=\{\bm{x}: 0.6\leq \norm{\bm{x}}_2\leq 0.8 \}$, the initial state set  $\mathcal{X}_0=\{\bm{x}:  \norm{\bm{x}}_2\leq 0.5 \}$, and the goal set  $\mathcal{X}_g=\{\bm{x}: \norm{\bm{x}-\bm{x}_{\text{g}}}_2\leq 0.2 \}$ with $\bm{x}_g=[\text{-}0.2,0]^\prime$. These regions are labeled in Fig. \ref{bicycle.1} using different colors. Fig. \ref{bicycle.1} shows the simulation of different trajectories of the  system under the learned policy, and Fig. \ref{bicycle.2} shows the learned Lyapunov-like and barrier certificate functions. We observe that these is no trajectory entering into the unsafe area, and all trajectories converge to the goal set.

\textbf{Planner UAV control.} \quad Finally, we test the method on a UAV flying in a planar. We define the system state variable as $\bm{x}=[x, y, \theta, \dot{x}, \dot{y},\dot{\theta}]^\prime$, where $(x,y)$ and $\theta$ represent the position and orientation of the UAV, respectively. Define the state space $\mathcal{X}=\{\bm{x}:\bm{x}_{\text{lb}}\leq \bm{x}\leq \bm{x}_{\text{ub}}\}$ with $\bm{x}_{\text{lb}}=[\text{-}1, \text{-}1,\text{-}1,\text{-}1]^\prime $ and $\bm{x}_{\text{ub}}=[1, 1,1,1]^\prime $, the unsafe state set $\mathcal{X}_u=\{\bm{x}: 0.9\leq \norm{\bm{x}}_2\leq 1 \}$, the initial state set  $\mathcal{X}_0=\{\bm{x}:  \norm{\bm{x}}_2\leq 0.5 \}$, and the goal state set  $\mathcal{X}_g=\{\bm{x}: \norm{\bm{x}}_2= 0 \}$. $\mathcal{X}_u$ and $\mathcal{X}_0$ are labeled in Fig. \ref{uav.1}. Fig. \ref{uav.1} shows the simulation of the state (norm) trajectories  of the  system under the learned policy, and Fig. \ref{uav.2} shows the learned Lyapunov-like and barrier certificate functions. From the results we can conclude that the learned policy is both safe and goal-reaching.

\section{Conclusions}
This paper proposes a method for learning provably safe and goal-reaching policies. The key to achieve so is to  learn the policy jointly with  learning two additional certificate functions: one is the barrier function, and the other is the Lyapunov-like function. We demonstrate the efficacy of the approach using different systems. We envision that this method provides a paradigm towards developing safe  and goal-oriented learning techniques. We also envision the benefits of the method to the control field, where we show that by taking advantage of learning techniques, general barrier and Lyapunov functions can be found using (deep) neural networks.

\section*{Broader Impact}

We envision that the results of this work could potentially benefit the learning community towards developing safe and goal-oriented learning algorithms. For example, in developing reinforcement learning algorithms, results of this work can be incorporated into the reinforcement learning framework to guarantee that the  policy search is constrained  within the set  of safe policies. For deep learning research, for examples, learning (deep) neural networks, neural ordinary differential equations, or  equilibrium models, use of the results of this work can guarantee that the learned model will provably converge towards to a goal set.
More importantly, inspired by the theoretical and experimental results of this work, we may conclude a more general learning paradigm: by including different certificate functions, one is able to simultaneously guarantee different aspects of performance of the learned models. This may benefit development of more complex but controllable learning systems in future.

We also envision that this work could benefit to the control field, because we provide a general solution on how to find the barrier and Lyapunov functions using (deep) neural networks.

The proposed methods in this work do not leverage any bias in data.

\bibliographystyle{unsrt}
\bibliography{safelearning} 

\newpage
\appendix

Appendices to the Neural Certificates for Safe Control Policies paper

\setcounter{equation}{0}
\renewcommand\theequation{I.\arabic{equation}}

\section{Proofs}\label{proofs}

Before presenting proofs for the theorems of this paper, we provide some preliminaries which will be used in deriving proofs. First, we present the definition of forward invariance of a set.
\begin{definition}[Forward invariance of a set]\label{forwardinvariance}
	Given a dynamical system $\dot{\bm{x}}=\bm{f}(\bm{x})$ with $\bm{f}:\mathcal{X}\rightarrow\mathcal{X}\subseteq\mathbb{R}^n$, we say a set $\mathcal{S}\subseteq\mathcal{X}$ is forward invariant if for every $\boldsymbol{x}(0)\in\mathcal{S}$, any state along the system trajectory $\bm{x}(t)$ starting from $\boldsymbol{x}(0)$ have $\boldsymbol{x}(t)\in\mathcal{S}$ for all $t\in \mathbb{R}^+_0$.
\end{definition}

Second, we provide the Nagumo’s Theorem \cite{blanchini1999set,blanchini2008set,ames2016control} which establishes a sufficient and necessary condition to verify the invariance of a sub-level set of a continuously differentiable  function.

\begin{theorem}[Nagumo’s Theorem \cite{blanchini1999set,blanchini2008set,ames2016control}]\label{nagumo}
	Consider a dynamical system $\dot{\bm{x}}=\bm{f}(\bm{x})$ with $\bm{f}:\mathcal{X}\rightarrow\mathcal{X}\subseteq\mathbb{R}^n$ and a continuously differentiable scalar function $h(\boldsymbol{x}):\mathcal{X}\rightarrow \mathbb{R}$. Define the zero sub-level set  of $h(\boldsymbol{x})$ as $\mathcal{C}=\{\bm{x}\in\mathcal{X}:h(\bm{x})\leq 0\}$. The following two conditions are equivalent:
	\begin{itemize}
		\item[1)] $\mathcal{C}$ is forward invariant in a sense of Definition \ref{forwardinvariance};
		\item[2)] $\nabla h(\bm{x})\bm{f}(\bm{x})\leq 0$,\,\, if \,\, $\bm{x}\in \{\bm{x}\in\mathcal{X}, h(\bm{x})=0\}$.
	\end{itemize}
\end{theorem}
The above Nagumo’s Theorem states that $\dot{h}(t)=\nabla h(\bm{x})\bm{f}(\bm{x})\leq 0$ on the boundary of the zero sub-level set $\mathcal{C}$ is a necessary and sufficient for $\mathcal{C}$ to be forward invariant. Please refer to  \cite{blanchini1999set,blanchini2008set,ames2016control} for more detail and proofs of this claim.

\subsection{Proof of Lemma \ref{theoremsafecontrol}} \label{proof.safe}

Assume that a barrier function $B(\bm{x})$ satisfying the three conditions in (\ref{cbf}) can be found. Take  any trajectory $\bm{x}_{\bm{u}}(t)$ in $\mathcal{X}$ that starts at some $\boldsymbol{x}(0)\in\mathcal{X}_0$ and consider the
evaluation of $B(\bm{x}_{\bm{u}}(t))$ along the trajectory. The condition (\ref{cbf.3}) directly indicates the second condition in the  Nagumo’s Theorem \ref{nagumo} holds, which is equivalent to say that  $\{\bm{x}\in\mathcal{X}: B(\bm{x})\leq 0\}$ is forward invariance according to the Nagumo’s Theorem. Thus, along the trajectory $\bm{x}_{\bm{u}}(t)$, $B(\bm{x}_{\bm{u}}(t))\leq 0$ holds for all $t\in \mathbb{R}^+_0$. Consequently, any such trajectory can never reach an unsafe state whose $B(\bm{x})$ is positive according to (\ref{cbf.2}).  We conclude that the safety of the system is guaranteed. \qed.

\subsection{Proof of Theorem \ref{theorm2}}
The proof of Theorem \ref{theorm2} consists of three steps.

First, we need to show that the following set
\begin{equation}\label{proof.goal.1}
\mathcal{A}=\{\bm{x}\in\mathcal{X}: V(\bm{x})\leq 0\}
\end{equation}
is closed and invariant. Its closeness is straightforward, and the invariance  can be proved by applying the  Nagumo’s Theorem \ref{nagumo}. Specifically, from the condition (\ref{equlyapu.2}), if $\bm{x}\in \{\bm{x}\in\mathcal{X}: V(\bm{x})= 0\}$, then $\nabla V(\bm{x})\bm{f}_{\bm{u}}(\bm{x})\leq 0$, which directly indicates the second condition in the  Nagumo’s Theorem \ref{nagumo} holds. Thus, we can say that $\mathcal{A}$ is invariant by applying  the  Nagumo’s Theorem \ref{nagumo}.

Second, we need to define another Lyapunov function
\begin{equation}\label{proof.goal.lyapunov}
V_{\mathcal{A}}(\boldsymbol{x})=
\begin{cases}
0 &\text{if $\bm{x}\in \mathcal{A}$}\\
V(\bm{x}) &\text{if $\bm{x} \in \mathcal{X}\textbackslash\mathcal{A}$}
\end{cases}.
\end{equation}
Combining the definition of the Lyapunov-like function in Definition \ref{deflyapunov}, it is easy to show the following properties of the Lyapunov function in (\ref{proof.goal.lyapunov}): (i) $V_{\mathcal{A}}(\boldsymbol{x})=0$ for all $\bm{x}\in\mathcal{A}$; (ii) $V_{\mathcal{A}}(\boldsymbol{x})>0$ for all $\bm{x} \in \mathcal{X}\textbackslash\mathcal{A}$ due to (\ref{proof.goal.1}); and (iii) for all $\bm{x} \in \mathcal{X}\textbackslash\mathcal{A}$, we have
\begin{equation}
V_{\mathcal{A}}(\boldsymbol{x})=V(\boldsymbol{x})\leq -\beta(V(\bm{x}))=-\beta(V_{\mathcal{A}}(\bm{x})),
\end{equation}
which is a result of directly applying the condition (\ref{equlyapu.2}) in Definition \ref{deflyapunov}.

Third, based on the results obtained in the first and second steps, and also from the fact that $V_{\mathcal{A}}(\bm{x})$ is continuous on its domain and continuously differentiable at every point $\bm{x} \in \mathcal{X}\textbackslash\mathcal{A}$, we can directly apply Theorem 2.8 in \cite{lin1996smooth} to show that  $\mathcal{A}$ is asymptotically stable; that is,  there exists a $\mathcal{KL}$-function $\gamma$ such that for any $\bm{x}(0) \in \mathcal{X}\textbackslash\mathcal{A}$, 
\begin{equation}
\norm{\bm{x}_{\bm{u}}(t)}_{\mathcal{A}}\leq\gamma(\norm{\bm{x}(0)}_{\mathcal{A}},t) 
\end{equation}
holds for  all  $t\in \mathbb{R}^{+}_0$.
Also combining  $\mathcal{A}\subseteq \mathcal{X}_g$ in (\ref{equlyapu.1}) of Definition \ref{deflyapunov}, it follows that 
\begin{equation}
\norm{\bm{x}_{\bm{u}}(t)}_{\mathcal{X}_g}\leq\norm{\bm{x}_{\bm{u}}(t)}_{\mathcal{A}}\leq\gamma(\norm{\bm{x}(0)}_{\mathcal{A}},t) 
\end{equation}
holds for  all  $t\in \mathbb{R}^{+}_0$. This directly indicates that the controlled system is goal-reaching by Definition \ref{defsatble}. Thus, we conclude that the existence of the Lyapunov-like function in Definition \ref{deflyapunov} guarantees the goal-reaching  of the controlled system. \qed.

\section{Experiment Details}\label{experiments}
\textbf{Pendulum system.}\quad
The equation of motion for the pendulum system is 
\begin{equation}\label{dyn.pendulum}
\ddot{\alpha}=-\frac{g}{l}\sin(\alpha)-\frac{d}{ml^2}\dot{\alpha}+\frac{u}{ml^2},
\end{equation}
with the constants set as $g=10, l=1, m=1, d=0.1$. We define the  state variable to be $\bm{x}=[\alpha,\dot{\alpha}]^\prime$. Define the state domain as $\mathcal{X}=\{\bm{x}:\bm{x}_{\text{lb}}\leq \bm{x}\leq \bm{x}_{\text{ub}}\}$ with $\bm{x}_{\text{lb}}=[-\pi, -5]^\prime $ and $\bm{x}_{\text{ub}}=[\pi, 5]^\prime $, the unsafe state set $\mathcal{X}_u=\{\bm{x}: 2.5\leq \norm{\bm{x}}_2\leq3 \}$, the goal state set  $\mathcal{X}_g=\{\bm{x}: \norm{\bm{x}}_2=0 \}$, and the initial state set  $\mathcal{X}_0=\{\bm{x}: \norm{\bm{x}}_2\leq 2 \}$.

We set the neural policy network to be one-layer linear function (without bias): $u=K\boldsymbol{x}$. For the neural barrier function $B_{\bm{\theta}}(\bm{x})$, we use a 2-16-16-1 fully connected network with $\tanh$ activation function. For the neural Lyapunov-like function $V_{\bm{\omega}}(\bm{x})$, we use a 2-16-16-1 fully connected network with $\tanh$ activation function, but the last layer is modified to be dot product operation. The learning rate is set as $10^{-3}$. Note that in our experiments, we always choose 4-layer neural networks for certificate functions, and the number of  nodes in layers is set as $n\text{-}8n\text{-}8n\text{-}1$ with $n$ is the  dimension of input layer (i.e., state dimension).

Only with Lyapunov-like certificate, the learned neural policy is
\begin{equation}
u=[-0.3286, -0.5950]\boldsymbol{x}.
\end{equation}
With both the Lyapunov-like and barrier  certificates, the learned neural policy is
\begin{equation}
u=[ 2.0120, -2.1343] \bm{x}.
\end{equation}

\textbf{Cartpole system.} The equation of the motion for the cartpole system is
\begin{subequations}
	\begin{align}
	\ddot{x}&=\frac{ u + m_p \sin\theta (l\dot{\theta}^2 - g \cos\theta)}{m_c + m_p (\sin\theta)^2},  \\
	\ddot{\theta}&=\frac{u\cos\theta+m_pl\dot{\theta}^2\cos\theta\sin\theta-(m_c+m_p)g*\sin\theta}{l(m_c + m_p (\sin\theta)^2)},
	\end{align}
\end{subequations}
with the constants set as $m_c=1, m_p=1, g=1, l=1$. The system state variable is defined as $\boldsymbol{x}=[x,\theta,\dot{x},\dot{\theta}]^\prime$, where $x$ is the position of the cart and $\theta$ is the angle between pole and upward direction.   Define the state space $\mathcal{X}=\{\bm{x}:\bm{x}_{\text{lb}}\leq \bm{x}\leq \bm{x}_{\text{ub}}\}$ with $\bm{x}_{\text{lb}}=[\text{-}1.3, \text{-}1.3, \text{-}1.3, \text{-}1.3]^\prime $ and $\bm{x}_{\text{ub}}=[1.3, 1.3, 1.3, 1.3]^\prime $, the unsafe state set $\mathcal{X}_u=\{\bm{x}: 0.9\leq \norm{\bm{x}}_2\leq 1.3 \}$, the goal state set  $\mathcal{X}_g=\{\bm{x}:  \norm{\bm{x}}_2=0 \}$, and the initial state set  $\mathcal{X}_0=\{\bm{x}: \norm{\bm{x}}_2\leq 0.8 \}$.

We set the neural policy network to be one-layer linear function (without bias): $u=K\boldsymbol{x}$. For the neural barrier function $B_{\bm{\theta}}(\bm{x})$, we use a 4-32-32-1 fully connected network with $\tanh$ activation function. For the neural Lyapunov-like function $V_{\bm{\omega}}(\bm{x})$, we use a 4-32-32-1 fully connected network with $\tanh$ activation function, but the last layer is modified to be dot product operation. The learning rate is set as $10^{-3}$.

Only with Lyapunov-like certificate, the learned neural policy is
\begin{equation}
u=[-0.0652, -0.2577, -1.3080, -0.6947]\boldsymbol{x}.
\end{equation}
With both the Lyapunov-like  and barrier certificates, the learned neural policy is
\begin{equation}
u=[-1.5064, -0.7969, -3.1892, -1.5950] \bm{x}.
\end{equation}

\textbf{Vehicle path tracking system.} The kinematic model of a wheeled vehicle tracking a reference path is given by \cite{chang2019neural}:
\begin{subequations}
	\begin{align}
	\dot{s}&=\frac{v\cos\theta_e)}{1-{d}_e\kappa(s)},\\
	\dot{d}_e&=v\sin(\theta_e),\\
	\dot{\theta}_e&=\frac{v\tan(u)}{L}-\frac{v\kappa(s)\cos\theta_e}{1-{d}_e\kappa(s)},
	\end{align}
\end{subequations}
where $\theta_e=\theta-\theta_r$ is the angle error between the  vehicle orientation $\theta$ and the reference path tangent angle $\theta_r$; $d_e$ is the distance error (see Figure 4. (c) in \cite{chang2019neural}); and the constants are $v=6$ and $L=1$. Assuming that the reference path is a unit circle. Define the state domain $\mathcal{X}=\{\bm{x}:\bm{x}_{\text{lb}}\leq \bm{x}\leq \bm{x}_{\text{ub}}\}$ with $\bm{x}_{\text{lb}}=[\text{-}0.8, \text{-}0.8]^\prime $ and $\bm{x}_{\text{ub}}=[0.8, 0.8]^\prime $, the  unsafe state set $\mathcal{X}_u=\{\bm{x}: 0.6\leq \norm{\bm{x}}_2\leq 0.8 \}$, the initial state set  $\mathcal{X}_0=\{\bm{x}:  \norm{\bm{x}}_2\leq 0.5 \}$, and the goal state set  $\mathcal{X}_g=\{\bm{x}: \norm{\bm{x}-\bm{x}_{\text{g}}}_2\leq 0.2 \}$ with $\bm{x}_g=[\text{-}0.2,0]^\prime$.

We set the neural policy network to be one-layer linear function (without bias): $u=K\boldsymbol{x}$. For the neural barrier function $B_{\bm{\theta}}(\bm{x})$, we use a 2-16-16-1 fully connected network with $\tanh$ activation function. For the neural Lyapunov-like function $V_{\bm{\omega}}(\bm{x})$, we use a 2-16-16-1 fully connected network with $\tanh$ activation function, but the last layer is modified to be dot product operation. The learning rate is set as $10^{-3}$. With both the Lyapunov-like and barrier certificates, the learned neural policy is
\begin{equation}
u=[-0.3662, -1.7802] \bm{x}.
\end{equation}

\textbf{UAV control.} \quad The motion of equation for a UAV flying in planar is given by:
\begin{subequations}
	\begin{align}
	\ddot{x}&=\frac{-(u_1 + u_2)\sin\theta}{m},\\
	\ddot{y}&=\frac{(u_1 + u_2)\cos\theta-mg}{m},\\
	\ddot{\theta}&=\frac{r(u_1-u_2)}{I},
	\end{align}
\end{subequations}
where the constants are set as $m=0.1, g=0.1, I=0.1, r=0.1$. We define the state variable as $\boldsymbol{x}=[x,y,\theta,\dot{x},\dot{y},\dot{\theta}]^\prime$ and the control variable $\boldsymbol{u}=[u_1, u_2]^\prime$.  Define the state domain $\mathcal{X}=\{\bm{x}:\bm{x}_{\text{lb}}\leq \bm{x}\leq \bm{x}_{\text{ub}}\}$ with $\bm{x}_{\text{lb}}=[\text{-}1, \text{-}1,\text{-}1,\text{-}1]^\prime $ and $\bm{x}_{\text{ub}}=[1, 1,1,1]^\prime $, the unsafe state set $\mathcal{X}_u=\{\bm{x}: 0.9\leq \norm{\bm{x}}_2\leq 1 \}$, the initial state set  $\mathcal{X}_0=\{\bm{x}:  \norm{\bm{x}}_2\leq 0.5 \}$, and the goal state set  $\mathcal{X}_g=\{\bm{x}: \norm{\bm{x}}_2= 0 \}$.

We set the neural policy network to be one-layer linear function (without bias): $\bm{u}=K\boldsymbol{x}$. For the neural barrier function $B_{\bm{\theta}}(\bm{x})$, we use a 6-48-48-1 fully connected network with $\tanh$ activation function. For the neural Lyapunov-like function $V_{\bm{\omega}}(\bm{x})$, we use a 6-48-48-1 fully connected network with $\tanh$ activation function,  but the last layer is modified as dot product operation. The learning rate is set as $10^{-3}$. With both the Lyapunov-like and barrier certificates, the learned neural policy is
\begin{equation}
\bm{u}=
\begin{bmatrix}
0.8185,&  0.8221,& -1.9815,&  2.4234, &-0.2271,& -1.8433\\
0.9136, &-1.0979,& -1.8189,& -0.0967, &-5.1917, & 0.3099
\end{bmatrix}
\bm{x}.
\end{equation}

\end{document}